\documentclass[prb,aps,amssymb,twocolumn,superscriptaddress,notitlepage]{revtex4-2}
\usepackage{amsmath}
\usepackage{amssymb}
\usepackage{amsthm}
\usepackage{amsfonts}
\usepackage{listings}
\usepackage{physics}
\usepackage{enumerate}
\usepackage{latexsym}
\usepackage{psfrag}
\usepackage{bm}
\usepackage{mathtools}
\usepackage{graphicx}
\usepackage[caption=false]{subfig}
\usepackage{blkarray}
\usepackage{array}
\usepackage{color}
\usepackage[makeroom]{cancel}
\usepackage[normalem]{ulem}
\usepackage{hyperref}
\hypersetup{
     colorlinks = true,
     linkcolor = magenta,
     citecolor = magenta
}

\def\beq{\begin{equation}}
\def\eeq{\end{equation}}
\def\bal{\begin{aligned}}
\def\eal{\end{aligned}}

\begin{document}

\title{Imaginary Time Formalism for Causal Nonlinear Response Functions}

\author{Sounak Sinha}
\affiliation{Department of Physics, University of Illinois Urbana-Champaign, Urbana IL 61801, USA}
\affiliation{Anthony J.~Leggett Institute for Condensed Matter Theory, University of Illinois Urbana-Champaign, Urbana IL 61801, USA}

\author{Barry Bradlyn}
\email{bbradlyn@illinois.edu}
\affiliation{Department of Physics, University of Illinois Urbana-Champaign, Urbana IL 61801, USA}
\affiliation{Anthony J.~Leggett Institute for Condensed Matter Theory, University of Illinois Urbana-Champaign, Urbana IL 61801, USA}

\date{\today}

\begin{abstract}
It is well established that causal linear response functions can be found by computing the much simpler imaginary time-ordered Matsubara functions and performing an analytic continuation. 
This principle is the basis for much of our understanding of linear response for interacting and disordered systems, via diagrammatic perturbation theory. 
Similar imaginary-time approaches have recently been introduced for computing nonlinear response functions as well, for example in [Annalen der Physik 536, 2300504 (2024); Physical Review X 11, 041006 (2021)], where the authors analytically continue the Matsubara functions to obtain the Keldysh response functions. 
In this work, we provide a proof of this connection to all orders in perturbation theory using an equation of motion based approach. We show by induction that causal nonlinear response functions at every order can be obtained from an analytic continuation of an appropriate time-ordered Matsubara function. 
We demonstrate this connection explicitly for second order response functions in the Lehmann representation.
As a byproduct of our approach, we derive an explicit expression for the Lehmann representation of $n$-th order response functions by solving the equations of motion.
We also use our result to find an analytic spectral density representation for both causal response functions and Matsubara functions.
As an example, we apply our method to derive the non-linear $A^3$ term in the $SU(2)$ spin Hall response of an insulator with spin rotation symmetry.
Finally, we show how our results lead to a family of generalized sum rules, focusing explicitly on the asymptotic expression for $n$-th harmonic generation rate. Our work opens the door to using imaginary time approaches to study nonlinear response functions in general condensed matter systems.
\end{abstract}
\maketitle

\section{Introduction}
Experiments on condensed matter systems involve perturbing systems with external time-dependent probe fields.
This pushes the system away from equilibrium, which induces a change in observables that can be measured.
An $n$-th order causal response function describes how an observable responds to a probe field at $n$-th order in perturbation theory.
They are of fundamental importance in condensed matter physics since  they are crucial in comparing experimental data with theory.
For instance, multi-dimensional spectroscopy allows direct measurement of nonlinear response functions in frequency space~\cite{mahmood2021observation, wan2019resolving, choi2020theory}. 
Additionally, topological materials have been shown to have approximately quantized and otherwise novel nonlinear optical conductivity which, if fully understood, could open the door to new applications in photovoltaics and information processing~\cite{alexandradinata2022quantization,de2017quantized}.
Causal response functions play a key role in understanding the optical response of such materials, and provide information on dynamics, symmetry and topology~\cite{gross1985local, patankar2018resonance, sodemann2015quantum,jankowski2024quantized, jankowski2025optical}.  
Previously, Refs.~\cite{parker2019diagrammatic,mckay2021optical,mckay2023spatially} provided a formalism for computing the non-linear optical conductivity by using a diagrammatic technique.
However, their technique produces the time-ordered, rather than the causal response functions.

Analytic solutions for the $n$-th order causal response function have been derived previously~\cite{jha1984nonlinear, watanabe2020generalized, watanabe2020general, ahn2022riemannian, sipe2000second,aversa1995nonlinear,lucarini2008response,passos2018nonlinear,peiponen1988nonlinear,peiponen1997dispersion,peiponen2004derivation,vartiainen1994meromorphic}, but cannot be easily applied to many-body systems because of the presence of nested commutators.
However, at linear order in perturbation theory,  linear response functions can be related to time-ordered two point correlation functions in imaginary time.
More specifically, the linear causal response function in frequency space can be analytically continued to the upper half plane, which can then be related to an imaginary time ordered (Matsubara) Green's function~\cite{matsubara1955new,fradkin2021quantum, abrikosov2012methods}.
The time ordering in the Matsubara Green's function allows one to use Feynman diagram techniques to compute the linear causal response functions for interacting Hamiltonians.
However, at higher orders in perturbation theory, the connection between imaginary time and real-time correlation functions is less clear. 
Though results exist for general Green's functions, the application of these results to response theory remains little explored. 
For instance, Ref. \cite{eliashberg1962transport} used the Lehmann representation to study the analytic properties of the four point correlation function of fermion operators on the complex plane and analytically continued the Matsubara correlation function to real frequencies, obtaining a finite conductivity of an interacting fermi gas.
Refs.~\cite{evans1992n, weldon2005search} considered general (unordered) real time correlation functions in frequency space and showed that only the fully causal and advanced linear combinations can be analytically extended to complex frequencies.
Refs. \cite{taylor1993real, guerin1994retarded, guerin1994rules} extended the results of Ref. \cite{evans1992n} to obtain all components of the Keldysh contour for Fermionic four point functions.
Refs.~\cite{kugler2021multipoint,ge2024analytic} investigated a partial spectral density representation to derive a method for analytically continuing any Matsubara Green's function to the causal Keldysh response function. 
Ref.~\cite{michishita2021effects} used similar methods to establish this result for second and third order response functions. 
However, these arguments are purely combinatorial, and do not make use of the physical properties satisfied by nonlinear response functions in particular. 

In this paper, we use an equations of motion technique to extend the imaginary time formalism to all orders in perturbation theory, and prove the relation between Matsubara and causal response functions for arbitrary interacting or disordered systems. 
Our method is more direct than that of Refs.~\cite{kugler2021multipoint, ge2024analytic, weldon2005search, evans1992n,michishita2021effects} since we only derive the connection between causal and Matsubara Green's functions, instead of analytically continuing all the out-of-time-order correlators that appear on the Keldysh contour. 
As such, we will be able to derive concrete formulas for the Lehmann representation of arbitrary nonlinear response and Matsubara functions (to our knowledge, Ref. \cite{kugler2021multipoint} is the only other work where the Lehmann kernel for arbitrary nonlinear response functions has been derived). We will also show how our approach can be used to derive families of generalized sum rules for nonlinear response functions. Our work puts on firm footing recent diagrammatic approaches for computing $n$-th order conductivities~\cite{parker2019diagrammatic,mckay2021optical,mckay2023spatially} and allows these approaches to generalize to interacting systems.

We first derive a recursion relation for the $n$-th order causal response function in frequency space in Sec.~\ref{sec:EOM}, which relates it to the $n-1$-th order causal response function by using the equations of motion for the density matrix.
We then show that there exists an imaginary time ordered $n$-point correlation function in frequency space that satisfies the same recursion relation.
Using the connection between first order causal response functions and the two-point Matsubara functions, in Sec.~\ref{sec:proof} we prove by induction the equivalence of the Lehmann representation of causal response functions and Matsubara functions to all orders in perturbation theory.
As a concrete example of the utility of our approach, we show in Sec.~\ref{sec:response} how our methods can be used to compute the non-Abelian (nonlinear) Chern-Simons response for insulators with spin rotation symmetry.
Finally, in Sec.~\ref{sec:spectral} we extend our proof to systems with multiple probe fields, and obtain a spectral density representation that yields both the causal and Matsubara functions even for systems with continuous spectra.
We then explicitly solve our recursion relation to derive a formula for $n$-th harmonic generation, as well as generalized sum rules for nonlinear response functions. We include full details of the proofs and derivations in Appendix \ref{Derivation of the recursion relation}. 

\section{Equations of Motion}\label{sec:EOM}
We begin with the Schrodinger equation for the density matrix $\rho(t)$, with dynamics governed by the time dependent Hamiltonian
\begin{equation}
    H(t)=H_0+e^{\epsilon t}f(t)B.
\end{equation}
The factor of $e^{\epsilon t}$ ensures that the perturbation, $f(t)$, goes to zero infinitely far in the past; the limit $\epsilon\rightarrow 0$ is taken at the end of all calculations.
We take the unperturbed density matrix $\rho(t\rightarrow-\infty)=\rho_0=\frac{1}{\mathcal{Z}}\exp(-\beta H_0)$ to be the thermal density matrix.
We expand $\rho(t)$ is powers of the perturbation $f(t)$,
\begin{flalign}
    \rho(t)=\sum_{n=0}^{\infty}\rho_n(t),
\end{flalign}
where $\rho_n(t)$ contains $n$ powers of the perturbation $f(t)$, and $\rho_0=\frac{1}{\mathcal{Z}}\exp(-\beta H_0)$.
With this expansion of $\rho(t)$, we see that the equation of motion for the $n$-th order contribution is given by
\begin{equation}
\label{EOM_for_rho}
    \dot{\rho_n}(t)=-i[H_0,\rho_n(t)]-ie^{\epsilon t}f(t)[B,\rho_{n-1}(t)].
\end{equation}
We are interested in the time evolution of an observable $A$ due to the perturbation $f(t)$.
The average $\langle A\rangle(t)$ can be expanded in powers of $f(t)$, defining the $n$-th order non-linear causal response function $\chi_{AB}^{(n)}$ as~\cite{jha1984nonlinear,kubo1957statistical},   
\begin{flalign}
\label{causal_response_function_def}
    &\mathrm{tr}(A\rho_n(t))\nonumber\\&\equiv e^{n\epsilon t}\int\prod_{i=1}^n\mathrm{d}t_if(t_i)e^{-\epsilon(t-t_i)}\chi_{AB}^{(n)}(t-t_1,\cdots,t-t_n).
\end{flalign}
Causality implies $\chi^{(n)}_{AB}$ vanishes whenever one of its arguments is negative.
Due to the exchange symmetry $t_i\leftrightarrow t_j$, only the completely symmetric part of $\chi^{(n)}_{AB}$ is experimentally observable. 
We Fourier transform the causal response function to frequency space and plug Eq.~\eqref{causal_response_function_def} into the equation of motion for $\rho_n(t)$, Eq.~\eqref{EOM_for_rho}.
This gives us a recurrence relation relating $\chi^{(n)}_{AB}(\omega_1,\cdots,\omega_n)$ to the $(n-1)$-th order causal response function,
\begin{flalign}
\label{recurrence_relation_for_causal_response}
    &\Big(\sum_{i=1}^n\omega^+_i\Big){\chi}^{(n)}_{AB}(\{\omega^+_i\}_{i=1}^n)\nonumber\\&={\chi}^{(n)}_{[A,H_0]B}(\{\omega^+_i\}_{i=1}^n)+\frac{{\chi}^{(n-1)}_{[A,B]B}(\{\omega^+_i\}_{i=2}^n)+(\omega^+_1\leftrightarrow\omega^+_{i\neq 1})}{n},
\end{flalign}
where $\omega_i^+=\omega_i+i\epsilon$.
This is a key result of this paper. 
Note that our Fourier transform conventions are,
\begin{flalign}
    \chi_{AB}^{(n)}(\{\omega_i\}_{i=1}^n)=\int\prod_{i=1}^n\left(\mathrm{d}t_ie^{i\omega_it_i}\right)\chi_{AB}^{(n)}(\{t_i\}_{i=1}^n).
\end{flalign}
We have used the explicit symmetry of $\chi^{(n)}$ under exchange of the $\omega$'s to symmetrize Eq.~\eqref{recurrence_relation_for_causal_response}. 
This will be crucial to match the symmetry properties of the Matsubara functions.
Additionally, if $[A,B]=0$, we will need to relate $\chi^{(n)}_{AB}$ to $\chi^{(n)}_{[A,H_0]B}$ using Eq.~\eqref{recurrence_relation_for_causal_response}. 
Then, if $[[A,H_0],B]\neq 0$, we use the recursion relation to reduce $\chi^{(n)}$ to $\chi^{(n-1)}$, and so on.

\section{Proof of Equivalence}\label{sec:proof}
Using the Lehmann representation, we expand $\chi^{(n)}_{AB}$ in powers of $B$ as
\begin{flalign}
\label{Lehmann_representation}
    &\chi^{(n)}_{AB}(\{\omega_i\}_{i=1}^n)\nonumber\\&=\frac{1}{\mathcal{Z}}\sum_{i_1,\cdots,i_{n+1}}B^{i_1}_{i_2}\cdots B^{i_n}_{i_{n+1}}A^{i_{n+1}}_{i_1}f^{(n)}_{i_1,\cdots,i_{n+1}}(\{\omega_i\}_{i=1}^n),
\end{flalign}
where $A^i_j=\langle i|A|j\rangle$, are the matrix elements of $A$ in the eigenbasis of $H_0$. 
In using the Lehmann representation, we are assuming that the causal response function can be written in terms of a spectral density that is a sum over Dirac delta functions.
This is always manifestly true for systems with discrete spectrum ~\cite{abrikosov2012methods}.
We will demonstrate later that the spectral density can be expressed as a sum of delta functions for all $n$ even for systems with continuous spectrum.
Plugging Eq.~\eqref{Lehmann_representation} into Eq.~\eqref{recurrence_relation_for_causal_response}, and exploiting the fact that Eq.~\eqref{recurrence_relation_for_causal_response} holds for arbitrary operators, we find that the $f^{(n)}$ satisfy
\begin{flalign}
\label{rec1_main}
    &f^{(n)}_{i_1,\cdots,i_{n+1}}(\{\omega_i\}_{i=1}^n)\nonumber\\&=\frac{f^{(n-1)}_{i_2,\cdots,i_{n+1}}(\{\omega_i\}_{i=2}^n)-f^{(n-1)}_{i_1,\cdots,i_{n}}(\{\omega_i\}_{i=2}^{n
    })+(\omega_1\leftrightarrow\omega_{i\neq 1})}{n(\omega_t^+-\epsilon_{i_1i_{n+1}})},
\end{flalign}
where $\omega_t^+=\sum_{i=1}^n\omega_i^+$ and $\epsilon_{ij}=\epsilon_i-\epsilon_j$, where $\epsilon_i$ are the eigenvalues of $H_0$.
Using the Kubo formula for linear response, we know that for $n=1$ we have~\cite{fradkin2013field}, 
\begin{equation}
    f^{(1)}_{i_1,i_2}(\omega)=\frac{e^{-\beta\epsilon_{i_2}}-e^{-\beta\epsilon_{i_1}}}{\omega^+-\epsilon_{i_1i_2}}.
\end{equation}
Combining with Eq.~\eqref{rec1_main}, we can evaluate the causal response for arbitrary $n$.

We now compare the causal response functions to the Matsubara Green's function,
\begin{equation}
\label{im_time_GF}
    \bar{\chi}^{(n)}_{AB}(\tau_1,\cdots,\tau_n)=\frac{1}{n!}\mathrm{tr}(\mathbb{T}A(0)B(-\tau_1)\cdots B(-\tau_n)\rho_0),
\end{equation}
where $\tau \in [0,\beta)$ and $B(\tau)=e^{\tau H_0}Be^{-\tau H_0}$.
Using the chain rule, we see that
\begin{flalign}
\label{chain_rule}
    \sum_{i=1}^n\frac{\partial}{\partial\tau_i}\bar{\chi}^{(n)}_{AB}(\tau_1,\cdots,\tau_n)&=\frac{\mathrm{d}}{\mathrm{d}\tau}\bar{\chi}^{(n)}_{AB}(\tau+\tau_1,\cdots,\tau+\tau_n)\Bigg|_{\tau=0}\nonumber\\&=-\bar{\chi}^{(n)}_{[A,H_0]B}(\tau_1,\cdots,\tau_n),
\end{flalign}
where we used the imaginary time translation invariance to move $\tau$ into the argument of $A$, which yields $[A,H_0]$ upon differentiation.
We next Fourier transform both sides of Eq.~\eqref{chain_rule} into (Matsubara) frequency space.
Integrating by parts, we see that in frequency space the correlation function satisfies a similar recurrence relation to Eq.~\eqref{recurrence_relation_for_causal_response}, namely,
\begin{flalign}
\label{recurrence_relation_for_Matsubara_function}
    &\Big(-i\sum_{i=1}^n\nu_i\Big)\bar{\chi}^{(n)}_{AB}(\{\nu_i\}_{i=1}^n)\nonumber\\&=-\bar{\chi}^{(n)}_{[A,H_0]B}(\{\nu_i\}_{i=1}^n)+\frac{\bar{\chi}^{(n-1)}_{[A,B]B}(\{\nu_i\}_{i=2}^n)+(\nu_1\leftrightarrow\nu_{i\neq 1})}{n}.
\end{flalign}
Here, the $\nu_i$s are Matsubara frequencies, which are constrained to be $\nu_i=\frac{2\pi}{\beta}m$ since $A$ and $B$ are (bosonic) observables~\cite{matsubara1955new,fradkin2013field}. 
We again use the Lehmann representation to expand $\bar{\chi}^{(n)}$ as a polynomial in matrix elements of $B$,
\begin{flalign}
\label{Lehmann_representation_Matsubara}
    &\bar{\chi}^{(n)}_{AB}(\{\nu_i\}_{i=1}^n)\nonumber\\&=\frac{1}{\mathcal{Z}}\sum_{i_1,\cdots,i_{n+1}}B^{i_1}_{i_2}\cdots B^{i_n}_{i_{n+1}}A^{i_{n+1}}_{i_1}g^{(n)}_{i_1,\cdots,i_{n+1}}(\{\nu_i\}_{i=1}^n).
\end{flalign}
Plugging this into Eq.~\eqref{recurrence_relation_for_Matsubara_function}, we see $g^{(n)}$ satisfies the same recursion relation as $-f^{(n)}$ with $\omega^+$ replaced by $i\nu$,
\begin{flalign}
    \label{rec2_main}
    &g^{(n)}_{i_1,\cdots,i_{n+1}}(\{\nu_i\}_{i=1}^n)\nonumber\\&=-\frac{g^{(n-1)}_{i_2,\cdots,i_{n+1}}(\{\nu_i\}_{i=2}^n)-g^{(n-1)}_{i_1,\cdots,i_{n}}(\{\nu_i\}_{i=2}^n)+(\nu_1\leftrightarrow\nu_{i\neq 1})}{n(i\nu_t-\epsilon_{i_1i_{n+1}})}.
\end{flalign}
Similar results for the Matsubara response were derived in \cite{rostami2021gauge}, but by evaluating the time ordered products directly.
Noting that, we have $\chi_{AB}^{(1)}(\omega)=-\bar{\chi}^{(1)}_{AB}(i\nu=\omega^+)$, we assume that 
\begin{flalign}
    &f^{(n-1)}_{i_1,\cdots,i_{n}}(\{\omega_i\}_{i=1}^{n-1})=(-1)^{(n-1)}g^{(n-1)}_{i_1,\cdots,i_{n}}(\{i\nu_i=\omega_i^+\}_{i=1}^{n-1}).
\end{flalign}
Finally, by comparing Eq.~\eqref{rec1_main} and \eqref{rec2_main}, we see our assumption implies $f^{(n)}$ continues to $g^{(n)}$ under $\omega_i^+\rightarrow i\nu_i$.  
Thus, to all orders in perturbation theory
\begin{equation}
\label{end_proof}
    \chi^{(n)}_{AB}(\{\omega^+_i\}_{i=1}^n)=(-1)^n\bar{\chi}^{(n)}_{AB}(\{i\nu_i=\omega^+_i\}_{i=1}^n),
\end{equation}
provided the continuation is carried out in the Lehmann representation. This concludes our proof.

We emphasize the simplicity of our method, which makes no assumptions about the underlying Hamiltonian.
In particular, Eq.~\eqref{end_proof} is true for Hamiltonians with many-body interactions and disorder, which are ubiquitous in most materials.
Additionally, our recursion relation Eq.~\eqref{rec1_main} allows us to construct the $f^{(n)}(\{\omega\})$ [and $g^{(n)}(\{i\nu\})$] for all $n$. 
In Appendix \ref{Direct calculation of the second order response function}, we explicitly demonstrate this for the second order causal response and the Matsubara three point functions.
Later we will derive a formula for $f^{(n)}(\{\omega\})$, which then gives us an exact formula for the Lehmann representation of $\chi^{(n)}_{AB}(\{\omega\})$ [and $\bar{\chi}^{(n)}_{AB}(\{i\nu\})$]. 
Finally, note that Eq.~\eqref{im_time_GF} is time ordered.
Therefore, we can use the diagrammatic techniques of Ref.~\cite{parker2019diagrammatic} to evaluate the right hand side of Eq.~\eqref{end_proof} to treat interactions or disorder perturbatively. 
Then, upon analytic continuation to real frequencies, we can obtain the causal optical response.
In Appendix \ref{Multiple perturbations}, we generalize this argument to Hamiltonians with multiple perturbations,
\begin{equation}
    H(t)=H_0+e^{\epsilon t}\sum_I f_I(t)B_I,
\end{equation}
showing that the causal response can still be related to the Matsubara Green's function.

From the recurrence relation of the Matsubara Green's function, Eq. \eqref{rec2_main}, we see that  if $i\nu_t=0$ and $\epsilon_{i_1 i_{n+1}}=0$, then there is an apparent divergence.
This divergence leads to ``anomalous'' contributions to the Matsubara Green's function; such contributions arise whenever there is are degeneracies in the spectrum of the Hamiltonian and a vanishing sum of Matsubara frequencies.
For instance, the first order susceptibility can be decomposed into, 
\begin{flalign}
    \bar\chi_{AB}^{(1)}(i\nu_n)=C_{AB}\beta\delta_{n,0}+[\bar\chi_{AB}^{(1)}(i\nu_n)]_{\mathrm{reg}},
\end{flalign}
where $C_{AB}$ is the anomalous part.
In the Lehmann representation, $C_{AB}$ has the following form, 
\begin{flalign}
    C_{AB}=\frac{1}{\mathcal{Z}}\sum_{\substack{nm,\\
    \epsilon_n=\epsilon_m}}e^{-\beta \epsilon_n}A^n_mB^m_n.
\end{flalign}
This should be contrasted with the regular part, which in the Lehmann representation is given by,
\begin{flalign}
    [\bar\chi_{AB}^{(1)}(i\nu)]_{\mathrm{reg}}=\frac{1}{\mathcal{Z}}\sum'_{nm}B^n_mA^m_n\frac{e^{-\beta \epsilon_m}-e^{-\beta\epsilon_n}}{i\nu-\epsilon_{mn}},
\end{flalign}
where a $'$ indicates that the degenerate subspace needs to be omitted from the sum.
Similarly, the second order susceptibility contains an anomalous term when $\nu_{12}=0$, 
\begin{flalign}
    [\bar\chi_{AB}(i\nu_1,i\nu_2)]_{\mathrm{anom}}=\delta(\nu_1+\nu_2)C_{AB}(i\nu_1),
\end{flalign}
where,
\begin{flalign}
    &C_{AB}(i\nu_1)\nonumber\\
    &=\frac{1}{\mathcal{Z}}\sum_{\substack{nl,\\
    \epsilon_n=\epsilon_l}}\sum_mB^n_mB^m_lA^l_n\left[\beta\frac{e^{-\beta\epsilon_l}}{i\nu_1-\epsilon_{lm}}+\frac{e^{-\beta\epsilon_m}-e^{-\beta\epsilon_l}}{(i\nu_1-\epsilon_{lm})^2}\right]\nonumber\\
    &+(\nu_1\rightarrow-\nu_1).
\end{flalign}
These anomalous contributions can be obtained by recomputing the Fourier transform of the imaginary time ordered correlators with $\nu_1=-\nu_2$, or equivalently, by setting $\nu_{12}=0$ in the explicit solution to the recurrence relation along with $\epsilon_i=\epsilon_j+\delta$, with infinitesimal $\delta$, and then taking the limit $\delta\rightarrow 0$ (see Refs \cite{halbinger2023spectral, ge2024analytic} for prescriptions to calculate the anomalous parts of arbitrary Matsubara correlators).
However, such anomalous contributions do not arise in the causal Green's function since $\omega^+=\omega+i\epsilon$ is always non zero. {The presence of anomalous terms reflects the difference between adiabatic and isothermal response, as we discuss further below.}
Thus, in order to have a well defined analytic continuation prescription, one needs to remove the anomalous contributions from the Matsubara Green's function \cite{watzenbock2022long, stevens1965green, kwok1969correlation}.
There are a number of methods to do so.
First, we can choose {to focus on the response of} operators $A$ and $B$ that are purely off diagonal in the degenerate subspace, $A^n_l=B^n_l=0$.
This is the case, for instance, in the calculation of the longitudinal conductivity, where one calculates the imaginary time ordered correlation function of the current operator $J(\mathbf{q})$. 
Keeping $\mathbf{q}\neq 0$ before the analytic continuation removes the anomalous terms in the Matsubara correlation function \cite{giuliani2008quantum}. 
Alternatively, we can remove the anomalous contributions directly from the spectral density, and analytically continue only the regular part of the Matsubara correlation functions, as was done in Refs. \cite{ge2024analytic, kugler2021multipoint}.
Ref. \cite{ge2024analytic} also demonstrates that analytic continuation of the anomalous terms in the Matsubara correlator yields Dirac $\delta$ contributions in the Keldysh path integral.
Our recurrence relation (along with the explicit solutions to it that we derive in the Appendix \ref{Explicit solution to the recursion relation}) continue to hold in the presence of the anomalous terms with the prescription that if $\epsilon_{ij}=0$ at zero frequency, we first set $\epsilon_{ij}=\delta$ and then take the limit of $\delta\rightarrow 0$.

Finally, we would like to point out that the $\omega\rightarrow 0$ limit of the causal response function actually gives us the adiabatic response of the system, rather than the isothermal susceptibility.
For a static (time-independent) perturbing field $f$, the static isothermal susceptibility is defined as \cite{giuliani2008quantum, mukerjee2008signatures, kubo1957statistical},
\begin{flalign} 
    \chi_{AB}^{\mathrm{iso}}=\lim_{f\rightarrow 0}\frac{\mathrm{tr}(A\rho_f)-\mathrm{tr}(A\rho_0)}{f},
\end{flalign}
where $\rho_f= e^{-\beta H_f}/\mathcal{Z}$.
Using the Lehmann representation, one can show,  \cite{giuliani2008quantum} 
\begin{flalign}
    \chi_{AB}^{\mathrm{iso}}=\lim_{\omega\rightarrow 0}\chi_{AB}(\omega)-\frac{\beta}{\mathcal{Z}}\sum_{\substack{nm,\\
    \epsilon_n=\epsilon_m}}e^{-\beta \epsilon_n}A^n_mB^m_n.
\end{flalign}
The right hand side of this equation is precisely equal to $\omega\rightarrow 0$ limit of $-\bar\chi_{AB}(i\nu_n\rightarrow\omega^+)$, without removal of the anomalous terms.
Thus the $\omega\rightarrow 0$ of the dynamical (adiabatic) susceptibility and the static isothermal susceptibility need not coincide.
The $\omega\rightarrow 0$ limit of the adiabatic susceptibility assumes that the thermal ensemble undergoes unitary evolution, and level crossings between orthogonal states are not allowed.
This assumption is true if the system is kept isolated from the rest of the world after preparation.
In contrast, the static isothermal susceptibility allows thermal interactions between the system and the environment, causing transitions between different orthogonal states.
We leave the derivation of the isothermal susceptibility from analytic continuation of imaginary time ordered correlators using our equation of motion based method to future work.

\section{Application to $SU(2)$ spin response}\label{sec:response}
We use our formalism to derive the Hall response of a spin rotation symmetric insulator.
Since the symmetry is non-Abelian, we expect the response to be given by the non-Abelian Chern Simons term,
\begin{flalign}
    S_{\mathrm{eff}}[A]=\frac{iN_3}{4\pi}\int\mathrm{Tr}\left(A\wedge\mathrm{d}A+\frac{2i}{3}A\wedge A\wedge A\right),
\end{flalign}
where the coefficient $N_3$ is a constant depending on the microscopic lattice Hamiltonian that we will derive.
Similar results for the effective action were obtained in Refs. \cite{qi2008topological, vayrynen2011soft, read2000paired, babu1987derivative} using Euclidean path integral methods.
The expression for $N_3$ can also be found in Refs.~\cite{gurarie2011single, ishikawa1986magnetic, volovik2003universe, ishikawa1987microscopic}.  
We derive the effective action directly from the more conventional causal response function, which we obtain by analytically continuing an imaginary time correlation function (hence the factor of $i$ in the $A^3$ term).
Since the connection between the causal and imaginary time response functions is not so well established at second order in perturbation theory, the $A^3$ term is a direct application of our result, and cannot be found in Refs. \cite{qi2008topological, vayrynen2011soft, read2000paired}.
We begin with the Hamiltonian,
\begin{flalign}
    H=\sum_{\mathbf{x},\mathbf{y}}c^{\dagger}_{\mathbf{x}\sigma}h(\mathbf{x}-\mathbf{y})c_{\mathbf{y}\sigma},
\end{flalign}
which is invariant under the symmetry $c_{\mathbf{x}}\rightarrow Uc_{\mathbf{x}}$, where $U=\exp(i\frac{\theta_a\sigma_a}{2})\in SU(2)$.
The Fermion operators satisfy the usual anti-commutation relations in momentum space, $\{c^{\dagger}_{\mathbf{q}},c_{\mathbf{p}}\}=\delta(\mathbf{q}-\mathbf{p})$, with all other anti-commutators vanishing.
We minimally couple the Hamiltonian to an $SU(2)$ gauge field $A=\frac{1}{2}A^a\sigma^a$,
\begin{flalign}
    H[A]=&-\frac{1}{2}\sum_{\mathbf{x}}A^a_0(t,\mathbf{x})c^{\dagger}_{\mathbf{x}}\sigma^ac_{\mathbf{x}}\nonumber\\
    &+\sum_{\mathbf{x},\mathbf{y}}c^{\dagger}_{\mathbf{x}}h(\mathbf{x}-\mathbf{y})\mathbb{P}e^{i\int_{\mathbf{y}}^{\mathbf{x}}\mathbf{A}(t,\mathbf{r})\cdot\mathrm{d}\mathbf{r}}c_{\mathbf{y}},
\end{flalign}
where $\mathbb{P}$ denotes path ordering.
In order to calculate the response, we need to expand the Hamiltonian up to cubic order in the gauge field $A$.
However, the cubic term only involves the spatial components of $A$ and cannot contribute to the $A^3$ term which has an $A^0$ in it.
Therefore, we only retain the paramagnetic and diamagnetic terms.
The Hamiltonian becomes,
\begin{flalign}
    H[A]=&H-\sum_{\mathbf{p}}A_{\mu}^a(t,-\mathbf{p})J^{\mu}_a(\mathbf{p})\nonumber\\
    &+\frac{1}{2}\sum_{\mathbf{p},\mathbf{q}}A_{\mu}^a(t,-\mathbf{p})A_{\nu}^b(t,-\mathbf{q})J^{\mu\nu}_{ab}(\mathbf{p},\mathbf{q}).
\end{flalign}
Our Fourier transform convention is, $c_{\mathbf{q},\omega}=\frac{1}{\sqrt{
V}}\sum_{\mathbf{x}}\int_{-\infty}^{\infty}\mathrm{d}te^{i\omega t-i\mathbf{p}\cdot\mathbf{x}}c(\mathbf{x},t)$, and $h(\mathbf{q})=\sum_{\mathbf{x}}e^{i\mathbf{q}\cdot\mathbf{x}}h(\mathbf{x})$, where $V$ is the sample volume.
Note that in the presence of orbitals that are not located at the unit cell origin one would need to dress the electron operators and the Bloch Hamiltonian with the embedding matrix, $V_{\alpha\beta}(\mathbf{k})=e^{i\mathbf{k}\cdot\mathbf{r}_{\alpha}}\delta_{\alpha\beta}$.
The paramagnetic and diamagnetic currents are given by \cite{mckay2023spatially}, 
\begin{flalign}
    &J^{\mu}_a(\mathbf{p})=\frac{1}{\sqrt{V}}\sum_{\mathbf{Q}}c^{\dagger}_{\mathbf{Q}-\mathbf{p}/2}\gamma^{\mu}_a(\mathbf{Q},\mathbf{p})c_{\mathbf{Q}+\mathbf{p}/2},\\ &J_{ab}^{\mu\nu}(\mathbf{p},\mathbf{q})=\frac{1}{V}\sum_{\mathbf{Q}}c^{\dagger}_{\mathbf{Q}-(\mathbf{p}+\mathbf{q})/2}\gamma^{\mu\nu}_{ab}(\mathbf{Q},\mathbf{p},\mathbf{q})c_{\mathbf{Q}+(\mathbf{p}+\mathbf{q})/2}.
\end{flalign}
We would like to emphasize that since $\mathbf{p},\mathbf{q}\neq 0$, we can ignore anomalous terms in the Matsubara frequency current correlation function, since there are no diagonal matrix elements in the energy basis.
The paramagnetic vertex is,
\begin{flalign}
    &\gamma^0_a(\mathbf{Q},\mathbf{p})=\frac{1}{2}\sigma_a,\\
    &\gamma^i_a(\mathbf{Q},\mathbf{p})=\frac{1}{2}\int_0^1\mathrm{d}\lambda\partial_ih(\mathbf{Q}+(\lambda-1/2)\mathbf{p})\sigma_a\sim\frac{1}{2}\partial_ih(\mathbf{Q})\sigma_a,
\end{flalign}
and the diamagnetic vertex is,
\begin{flalign}
    &\gamma^{00}_{ab}(\mathbf{Q},\mathbf{p},\mathbf{q})=\gamma^{0i}_{ab}(\mathbf{Q},\mathbf{p},\mathbf{q})=\gamma^{i0}_{ab}(\mathbf{Q},\mathbf{p},\mathbf{q})=0,\nonumber\\&\gamma^{ij}_{ab}(\mathbf{Q},\mathbf{p},\mathbf{q})\nonumber=\frac{1}{4}\partial_i\partial_jh(\mathbf{Q})\delta_{ab}.
\end{flalign}
In both the vertices we have approximated the integrals over $\lambda$ by the midpoint rule, and we also used $\Theta(0)=\frac{1}{2}$.
The cubic term, which is of the form $c^{\dagger}\partial^3h\sigma c$, does not contribute to the response since the expectation value gives us $\mathrm{Tr}\partial^3h \sigma$, which is zero since the Hamiltonian is independent of spin and Pauli matrices are traceless.
The response of the system is given by the expectation value of the current operator in the presence of the gauge fields,
\begin{flalign}
    \langle J^{\mu}_a(\mathbf{p},t)\rangle&=\frac{1}{\mathcal{Z}}\mathrm{tr}\left(J^{\mu}_a(\mathbf{p})e^{-\beta H}\right)\nonumber\\&+\mathrm{tr}(J^{\mu}_a(\mathbf{p})\rho_1(t))+\mathrm{tr}(J^{\mu}_a(\mathbf{p})\rho_2(t)).
\end{flalign}
The first term is momentum independent and contributes to the diamagnetic response, which is zero for insulators.
We therefore begin with the second term,
\begin{flalign}
    &\mathrm{tr}(J_a^{\mu}(\mathbf{p})\rho_1(t))&\nonumber\\&=-\int\frac{\mathrm{d}\omega}{2\pi} e^{-i\omega^+ t}\sum_{\mathbf{q}} \chi^{(1)}_{J_a^{\mu}(\mathbf{p})J_b^{\nu}(\mathbf{q})}(\omega^+)A^b_{\nu}(\omega,-\mathbf{q})+\mathcal{O}(A^2)\nonumber\\&=\int\frac{\mathrm{d}\omega}{2\pi} e^{-i\omega^+ t}\sum_{\mathbf{q}} \bar{\chi}^{(1)}_{J_a^{\mu}(\mathbf{p})J_b^{\nu}(\mathbf{q})}(i\nu_B=\omega^+)A^b_{\nu}(\omega,-\mathbf{q})+\mathcal{O}(A^2)\nonumber\\&=\int\frac{\mathrm{d}\omega}{2\pi} e^{-i\omega^+ t}\sum_{\mathbf{q}}\Pi^{\mu\nu}_{ab}(\mathbf{p},\mathbf{q},i\nu_B=\omega^+)A^b_{\nu}(\omega,-\mathbf{q})+\mathcal{O}(A^2),
\end{flalign}
where $\nu_B$ is a Bosonic Matsubara frequency.
In the second line we used our formula to convert the first order causal response function into an imaginary time ordered correlation function. 
Note that the $\mathcal{O}(A^2)$ term involves a correlation function of the paramagnetic and the diamagnetic current, and therefore can contribute to the $A^3$ term in the response.
However, since the paramagnetic term has a Pauli matrix in it, the trace over the spin indices gives zero. 
Hence the $\mathcal{O}(A^2)$ term vanishes.
Calculating $\Pi^{\mu\nu}_{ab}$ involves the two point correlation function of the currents in momentum space,
\begin{widetext}
\begin{flalign}
    &\Pi^{\mu\nu}_{ab}(\mathbf{p},\mathbf{q},i\nu_B)\nonumber\\&=\frac{1}{\mathcal{Z}}\int_0^{\beta}\mathrm{d}\tau e^{i\nu_B\tau}\mathrm{tr}(J^{\mu}_a(\mathbf{p})J_b^{\nu}(\mathbf{q},-\tau)e^{-\beta H})\nonumber\\&=\frac{1}{V}\int_0^{\beta}\mathrm{d}\tau e^{i\nu_B\tau}\sum_{\mathbf{Q}\mathbf{Q}'}[\gamma^{\mu}_a(\mathbf{Q},\mathbf{p})]_{\alpha\beta}[\gamma^{\nu}_b(\mathbf{Q}',\mathbf{q})]_{\gamma\delta}\left\langle c^{\dagger}_{\mathbf{Q}-\frac{\mathbf{p}}{2}\alpha}(\tau)c_{\mathbf{Q}'+\frac{\mathbf{q}}{2}\delta}\right\rangle\left\langle c_{\mathbf{Q}+\frac{\mathbf{p}}{2}\beta}(\tau)c^{\dagger}_{\mathbf{Q}'-\frac{\mathbf{q}}{2}\gamma}\right\rangle\nonumber\\&=\frac{1}{V}\int_0^{\beta}\mathrm{d}\tau e^{i\nu_B\tau}\sum_{\mathbf{Q}\mathbf{Q}'}[\gamma^{\mu}_a(\mathbf{Q},\mathbf{p})]_{\alpha\beta}[\gamma^{\nu}_b(\mathbf{Q}',\mathbf{q})]_{\gamma\delta}\delta(\mathbf{Q}-\mathbf{Q}')\delta(\mathbf{p}+\mathbf{q})\left\langle c^{\dagger}_{\mathbf{Q}-\frac{\mathbf{p}}{2}\alpha}(\tau)c_{\mathbf{Q}'+\frac{\mathbf{q}}{2}\delta}\right\rangle\left\langle c_{\mathbf{Q}+\frac{\mathbf{p}}{2}\beta}(\tau)c^{\dagger}_{\mathbf{Q}'-\frac{\mathbf{q}}{2}\gamma}\right\rangle,
\end{flalign}
\end{widetext}
where we used Wick's theorem.
Using the Euclidean time Green's function for the Fermion operators, given by ($\nu_1,\nu_2$ are Fermionic Matsubara frequencies),
\begin{flalign}
\label{Green's_function}
    &G^{(1)}_{\beta\alpha}(\mathbf{p},i\nu_1;\mathbf{q},i\nu_2)=\int_0^{\beta}\mathrm{d}^2\tau e^{i(-\nu_1\tau_1+\nu_2\tau_2)}\langle \mathbb{T}c^{\dagger}_{\mathbf{q}\alpha}(\tau_1)c_{\mathbf{p}\beta}(\tau_2)\rangle\nonumber\\
    &=-\delta(\mathbf{p}-\mathbf{q})\delta(\nu_1-\nu_2)\Big[\Big\{i\nu_2+h(\mathbf{q})\Big\}^{-1}\Big]_{\beta\alpha}\nonumber\\&=\delta(\mathbf{p}-\mathbf{q})\delta(\nu_1-\nu_2)G^{(1)}_{\beta\alpha}(\mathbf{q},i\nu_2),\nonumber\\
    &
    G^{(2)}_{\alpha\beta}(\mathbf{p},i\nu_1;\mathbf{q},i\nu_2)=\int_0^{\beta}\mathrm{d}^2\tau e^{i(\nu_1\tau_1-
    \nu_2\tau_2)}\langle \mathbb{T}c_{\mathbf{q}\alpha}(\tau_1)c^{\dagger}_{\mathbf{p}\beta}(\tau_2)\rangle\nonumber\\
    &=-\delta(\mathbf{p}-\mathbf{q})\delta(\nu_1-\nu_2)\Big[\Big\{i\nu_2-h(\mathbf{q})\Big\}^{-1}\Big]_{\alpha\beta}\nonumber\\&=\delta(\mathbf{p}-\mathbf{q})\delta(\nu_1-\nu_2)G^{(2)}_{\alpha\beta}(\mathbf{q},i\nu_2),
\end{flalign}
we have,
\begin{flalign}
    &\Pi^{\mu\nu}_{ab}(\mathbf{p},\mathbf{q},i\nu_B)=\delta(\mathbf{p}+\mathbf{q})\Pi_{ab}^{\mu\nu}(\mathbf{p},i\nu_B),
\end{flalign}
where, defining $\mathbf{Q}_{\pm}=\mathbf{Q}\pm\mathbf{p}/2$ and $\nu_{F-}'=\nu_F'-\nu_B$,
\begin{flalign}
    &\Pi_{ab}^{\mu\nu}(\mathbf{p},i\nu_B)\nonumber\\
    &=\sum_{\mathbf{Q},i\nu_F'}\mathrm{Tr}[\gamma^{\mu}_a(\mathbf{Q},\mathbf{p})]G^{(2)}(\mathbf{Q}_+,i\nu_F')[\gamma^{\nu}_b(\mathbf{Q},-\mathbf{p})]G^{(2)}(\mathbf{Q}_-,i\nu_{F-}')
\end{flalign}
Note that, in deriving the Green's function, we used the equations of motion for the Fermion operators,
\begin{flalign}
    &e^{\tau H}c_{\mathbf{q}\alpha}e^{-\tau H}=\sum_{\beta}\Big[e^{-\tau h(\mathbf{q})}\Big]_{\alpha\beta}c_{\mathbf{q}\beta},\nonumber\\
    &e^{\tau H}c^{\dagger}_{\mathbf{q}\alpha}e^{-\tau H}=\sum_{\beta}c^{\dagger}_{\mathbf{q}\beta}\Big[e^{\tau h(\mathbf{q})}\Big]_{\beta\alpha}.
\end{flalign}
To get the $A\mathrm{d}A$ term, we need to expand this function out to linear order in the momentum.
As mentioned before, the momentum independent contribution vanishes for an insulator.
Removing the $(2)$ from the Green's function, the linear order term is given by,
\begin{widetext}
\begin{flalign}
\label{photon_prop}
    \Pi^{\mu\nu}_{ab}(\mathbf{p},i\nu_B)&\sim-\frac{1}{2\beta V}\delta_{ab}\sum_{\mathbf{Q},i\nu'_F}\Big[\mathrm{Tr}\Big(\partial_{\mu}[G^{}(\mathbf{Q},i\nu'_F)]^{-1}\partial_iG^{}(\mathbf{Q},i\nu'_F)\partial_{\nu}[G^{}(\mathbf{Q},i\nu'_F)]^{-1}G^{}(\mathbf{Q},i\nu'_F)\Big)\frac{p^i}{2}\nonumber\\&\hspace{2.5cm}-\mathrm{Tr}\Big(\partial_{\mu}[G^{}(\mathbf{Q},i\nu'_F)]^{-1}G^{}(\mathbf{Q},i\nu'_F)\partial_{\nu}[G^{}(\mathbf{Q},i\nu'_F)]^{-1}\partial_iG^{}(\mathbf{Q},i\nu'_F)\Big)\frac{p^i}{2}\nonumber\\&\hspace{2.5cm}+\mathrm{Tr}\Big(\partial_{\mu}[G^{}(\mathbf{Q},i\nu'_F)]^{-1}G^{}(\mathbf{Q},i\nu'_F)\partial_{\nu}[G^{}(\mathbf{Q},i\nu'_F)]^{-1}\partial_0G^{}(\mathbf{Q},i\nu'_F)\Big)(i\nu_B)\Big]\nonumber\\
    &=\frac{1}{2\beta V}\sum_{\mathbf{Q},i\nu_F'}\Big[\mathrm{Tr}\Big(\frac{1}{2}\Big\{[[G^{-1}\partial_{\mu}G],[G^{-1}\partial_{\nu}G]]\Big\}(G^{-1}\partial_iG)\Big)p^i+\mathrm{Tr}\Big([G^{-1}\partial_{\mu}G][G^{-1}\partial_{\nu}G]G^{-1}\partial_0G\Big)(-i\nu_B)\Big].
\end{flalign}
\end{widetext}
In deriving this expression, we have used $\gamma^{\mu}_a(\mathbf{Q})=\partial_{\mu}[G^(\mathbf{Q})]^{-1}\frac{\sigma^a}{2}$, where $\partial_{\mu}=\frac{\partial}{\partial p_\mu}=\left(\frac{\partial}{\partial(-i\nu)},\frac{\partial}{\partial p_i}\right)$.
We also assumed that the states of the system can be written as a tensor product of eigenstates of the Bloch Hamiltonian and the spin matrices, $|U_m(\mathbf{k})\rangle\otimes|\sigma\rangle$, and the Green's functions do not mix spin and band indices. This allowed us to trace over the Pauli matrices $\sigma_a\sigma_b$ to obtain $2\delta_{ab}$ in Eq. \eqref{photon_prop}.
The current can be thought of as a part of an effective action,
\begin{flalign}
    \langle J^{\mu}_a(\mathbf{p},t)\rangle=\frac{\delta S_{\mathrm{eff}}[A]}{\delta A^a_{\mu}(-\mathbf{p},t)},
\end{flalign}
from which we see,
\begin{flalign}
    &S_{\mathrm{eff}}[A]\nonumber\\&=\frac{1}{2}\int\mathrm{d}t\sum_{\mathbf{p}}A^a_{\mu}(-\mathbf{p},t)\int\frac{\mathrm{d}\omega}{2\pi}e^{-i\omega^+t}\Pi_{ab}^{\mu\nu}(\mathbf{p},\omega)A^b_{\nu}(\omega,\mathbf{p})\nonumber\\&=\frac{1}{2}\int\frac{\mathrm{d}\omega}{2\pi}\sum_{\mathbf{p}}A_{\mu}(-\mathbf{p},-\omega)\Pi_{ab}^{\mu\nu}(\mathbf{p},\omega)A_{\nu}(\mathbf{p},\omega),
\end{flalign}
where we have taken the $\epsilon\rightarrow 0$ limit.
If we look at the second term in the expression for $\Pi_{ab}^{\mu\nu}(\mathbf{p},\omega)$ (the term proportional to $\omega$), using the effective action, we can replace the trace by $\mathrm{Tr}\Big(\frac{1}{2}\Big\{[[G^{-1}\partial_{\mu}G],[G^{-1}\partial_{\nu}G]]\Big\}G^{-1}\partial_0G\Big)$.
Therefore, the effective action is ($p_{\lambda}=(-\omega,\mathbf{p})$),
\begin{flalign}
\label{effective_action}
    S_{\mathrm{eff}}[A]=\frac{1}{4}\int\frac{\mathrm{d}\omega}{2\pi}\sum_{\mathbf{p}}A_{\mu}(-\mathbf{p},-\omega)h^{\mu\nu\lambda}p_{\lambda}A_{\nu}(\mathbf{p},\omega)+\mathcal{O}(A^3),
\end{flalign}
where 
\begin{equation}
\label{photon_self_energy}
    h^{\mu\nu\lambda}=\frac{1}{2\beta V}\sum_{\mathbf{Q},i\nu'_F}\mathrm{Tr}\Big([[G^{-1}\partial_{\mu}G],[G^{-1}\partial_{\nu}G]](G^{-1}\partial_{\lambda}G)\Big).
\end{equation}
Now, using the fact that 
\begin{equation}
    c^{\mu\nu\lambda}=\frac{1}{2\beta V}\sum_{\mathbf{Q},i\nu'_F}\Big[\mathrm{Tr}\Big([G^{-1}\partial_{\mu}G][G^{-1}\partial_{\nu}G][G^{-1}\partial_{\lambda}G\Big)\Big],
\end{equation}
satisfies $c^{\mu\nu\lambda}=c^{\nu\lambda\mu}=c^{\lambda\mu\nu}$, we convert Eq. \eqref{photon_self_energy} to,
\begin{flalign}
    &\frac{1}{2\beta V}\sum_{\mathbf{Q},i\nu'_F}\mathrm{Tr}\Big([[G^{-1}\partial_{\mu}G],[G^{-1}\partial_{\nu}G]](G^{-1}\partial_{\lambda}G)\Big)\nonumber\\&=\sum_{\mathbf{Q},i\nu'_F}\frac{\epsilon^{\alpha\beta\gamma}}{6\beta V}\mathrm{Tr}\Big([G^{-1}\partial_{\alpha}G][G^{-1}\partial_{\beta}G][G^{-1}\partial_{\gamma}G]\Big)\times\epsilon^{\mu\nu\lambda}.
\end{flalign}
This allows us to write the effective action Eq. \eqref{effective_action} as,
\begin{flalign}
    S_{\mathrm{eff}}[A]&=\frac{N_3}{8\pi}\int\frac{\mathrm{d}\omega}{2\pi}\sum_{\mathbf{p}}A^a_{\mu}(-\mathbf{p},-\omega)\epsilon^{\mu\nu\lambda}(p_{\lambda})A^a_{\nu}(\mathbf{p},\omega)\nonumber\\&=\frac{N_3}{8\pi}\int\frac{\mathrm{d}\omega}{2\pi}\sum_{\mathbf{p}}A^a_{\mu}(-\mathbf{p},-\omega)\epsilon^{\mu\nu\lambda}(-p_{\nu})A^a_{\lambda}(\mathbf{p},\omega)\nonumber\\&=i\frac{N_3}{8\pi}\int\frac{\mathrm{d}\omega}{2\pi}\sum_{\mathbf{p}}A^a_{\mu}(-\mathbf{p},-\omega)\epsilon^{\mu\nu\lambda}(ip_{\nu})A^a_{\lambda}(\mathbf{p},\omega)\nonumber\\&=i\frac{N_3}{8\pi}\int A^a\wedge\mathrm{d}A^a,
\end{flalign}
where $N_3$ is purely imaginary ($N_3^*=-N_3$), and is given by,
\begin{flalign}
\label{N_3_def}
    N_3&=-\frac{2\pi}{\beta V}\frac{\epsilon^{\mu\nu\lambda}}{6}\sum_{\mathbf{Q},i\nu_F}\mathrm{Tr}\Big([G^{-1}\partial_{\mu}G][G^{-1}\partial_{\nu}G][G^{-1}\partial_{\lambda}G]\Big).
\end{flalign}
Next, we calculate the second order term in the current response by converting the causal response function into an imaginary time ordered correlation function.
Unlike previous approaches which either work entirely in imaginary time~\cite{qi2008topological} or deduce the nonlinear response indirectly from gauge invariance constraints~\cite{read2000paired}, our approach allows us to directly compute the causal nonlinear spin response function.
We have,
\begin{widetext}
\begin{flalign}
    &\mathrm{tr}(J^{\mu}_a(\mathbf{p})\rho_2(t))\nonumber\\
    &=\int\mathrm{d}\omega\sum_{\mathbf{q}_1,\mathbf{q}_2}\int\frac{\mathrm{d}\omega_1\mathrm{d}\omega_2}{(2\pi)^2}e^{-i\omega_1^+t-i\omega_2^+t}A^b_{\nu}(-\mathbf{q}_1,\omega_1)A^c_{\lambda}(-\mathbf{q}_2,\omega_2)\delta(\omega-\omega_1-\omega_2)\chi^{(2)}_{J^{\mu}_a(\mathbf{p}) J^{\nu}_b(\mathbf{q}_1)J^{\lambda}_c(\mathbf{q}_1)}(\omega_1^+,\omega_2^+)\nonumber\\
    &=\int\mathrm{d}\omega\sum_{\mathbf{q}_1,\mathbf{q}_2}\int\frac{\mathrm{d}\omega_1\mathrm{d}\omega_2}{(2\pi)^2}e^{-i\omega_1^+t-i\omega_2^+t}A^b_{\nu}(-\mathbf{q}_1,\omega_1)A^c_{\lambda}(-\mathbf{q}_2,\omega_2)\delta(\omega-\omega_1-\omega_2)\Pi_{abc}^{\mu\nu\lambda}(\mathbf{p},\mathbf{q}_1,\mathbf{q}_2;i\nu_1=\omega_1^+,i\nu_2=\omega_2^+).
\end{flalign}
Using Wick's theorem, $\Pi^{\mu\nu\lambda}_{abc}$ evaluates to,
\begin{flalign}
    \Pi^{\mu\nu\lambda}_{abc}(\mathbf{p},\mathbf{q}_1,\mathbf{q}_2;i\nu_1,i\nu_2)\sim\delta(\mathbf{p}+\mathbf{q}_1+\mathbf{q}_2)\frac{1}{2 V^{3/2}}\int\mathrm{d}^2\tau& \sum_{\mathbf{Q}}[\gamma^{\mu}_a(\mathbf{Q})]_{\alpha_1\alpha_2}[\gamma^{\nu}_b(\mathbf{Q})]_{\beta_1\beta_2}[\gamma^{\lambda}_c(\mathbf{Q})]_{\gamma_1\gamma_2}\nonumber\\
    &\times[\langle c^{\dagger}_{\mathbf{Q}\alpha_1}(\tau_1)c_{\mathbf{Q}\beta_2}\rangle\langle c_{\mathbf{Q}\alpha_2}(\tau_2)c_{\mathbf{Q}\gamma_1}^{\dagger}\rangle\langle \mathbb{T}c^{\dagger}_{\mathbf{Q}
    \beta_1}(\tau_2)c_{\mathbf{Q}\gamma_2}(\tau_1)\rangle\nonumber\\
    &+\langle c^{\dagger}_{\mathbf{Q}\alpha_1}(\tau_2)c_{\mathbf{Q}\gamma_2}\rangle\langle c_{\mathbf{Q}\alpha_2}(\tau_1)c^{\dagger}_{\mathbf{Q}\beta_1}\rangle\langle\mathbb{T}c_{\mathbf{Q}\beta_2}(\tau_2)c^{\dagger}_{\mathbf{Q}\gamma_1}(\tau_1)\rangle].
\end{flalign}
\end{widetext}
In the last line, we set the external momenta and frequency to zero.
At this point, we would like to highlight that Eq. \eqref{rec2_main} from the main text greatly simplifies this calculation by allowing us to directly relate the causal and the Matsubara Green's function without having to resort to the Keldysh contour methods of Ref. \cite{ge2024analytic,kugler2021multipoint}.
Since the Matsubara Green's functions are (imaginary) time ordered, we were able to use Wick's theorem to simplify the six-point Fermion correlation function.
Now, using the definition of the Green's functions, Eq. \eqref{Green's_function}, we have,
\begin{widetext}
\begin{flalign}
    &\Pi^{\mu\nu\lambda}_{abc}(\mathbf{p},\mathbf{q}_1,\mathbf{q}_2;i\nu_1,i\nu_2)\nonumber\\
    &\sim\delta(\mathbf{p}+\mathbf{q}_1+\mathbf{q}_2)\frac{1}{2 V^{3/2}}\int\mathrm{d}^2\tau \sum_{\mathbf{Q}}[\gamma^{\mu}_a(\mathbf{Q})]_{\alpha_1\alpha_2}[\gamma^{\nu}_b(\mathbf{Q})]_{\beta_1\beta_2}[\gamma^{\lambda}_c(\mathbf{Q})]_{\gamma_1\gamma_2}\nonumber\\
    &\hspace{3cm}\times[G^{(1)}_{\beta_2\alpha_1}(\mathbf{Q},\tau_1)G_{\alpha_2\gamma_1}^{(2)}(\mathbf{Q},\tau_2)G^{(1)}_{\gamma_2\beta_1}(\mathbf{Q},\tau_2-\tau_1)+G^{(1)}_{\gamma_2\alpha_1}(\mathbf{Q},\tau_2)G^{(2)}_{\alpha_2\beta_1}(\mathbf{Q},\tau_1)G^{(2)}_{\beta_2\gamma_1}(\mathbf{Q},\tau_2-\tau_1)]\nonumber\\
    &=\delta(\mathbf{p}+\mathbf{q}_1+\mathbf{q}_2)\frac{1}{2 \beta 
    V^{3/2}}\nonumber\\
    &\times\sum_{\mathbf{Q},i\nu'_F}[\mathrm{Tr}(\gamma^{\nu}_b(\mathbf{Q})G(\mathbf{Q},i\nu'_F)\gamma^{\mu}_a(\mathbf{Q})G(\mathbf{Q},i\nu_F')\gamma^{\lambda}_c(\mathbf{Q})G(\mathbf{Q},i\nu_F'))-\mathrm{Tr}(\gamma^{\lambda}_c(\mathbf{Q})G(\mathbf{Q},i\nu_F')\gamma^{\mu}_a(\mathbf{Q})G(\mathbf{Q},i\nu_F')\gamma^{\nu}_b(\mathbf{Q})G(\mathbf{Q},i\nu'_F))].
\end{flalign}
\end{widetext}
To simplify the last expression, we note that the spin vertices are proportional to $\sigma_a$, the trace completely anti-symmetrizes the products of the vertices.
Next, using $\gamma^{\mu}_a(\mathbf{Q})=\partial_{\mu}G^{-1}(\mathbf{Q})\frac{\sigma_a}{2}$, we get,
\begin{flalign}
    &-i4\beta V^{3/2}\Pi^{\mu\nu\lambda}_{abc}(\mathbf{p},\mathbf{q}_1,\mathbf{q}_2;i\nu_1,i\nu_2)&\nonumber\\&\sim -\delta(\mathbf{p}+\mathbf{q}_1+\mathbf{q}_2)\epsilon_{abc}\sum_{\mathbf{Q},i\nu_F'}\mathrm{Tr}(G\partial_{\mu}G^{-1}G\partial_{\nu}G^{-1}G\partial_{\lambda}G^{-1})\nonumber\\
    &=\delta(\mathbf{p}+\mathbf{q}_1+\mathbf{q}_2)\epsilon_{abc}\sum_{\mathbf{Q},i\nu_F'}\mathrm{Tr}(G^{-1}\partial_{\mu}GG^{-1}\partial_{\nu}GG^{-1}\partial_{\lambda}G),
\end{flalign}
where we used $\mathrm{Tr}_s(\sigma_a\sigma_b\sigma_c)=2i\epsilon_{abc}$.
Next, we note that the epsilon tensor completely anti-symmetrizes the product of gauge fields, which ends up anti-symmetrizing the trace of the propagators.
This allows us to write,
\begin{flalign}
    \Pi^{\mu\nu\lambda}_{abc}(\mathbf{p},\mathbf{q}_1,\mathbf{q}_2;i\nu_1,i\nu_2)=-\delta(\mathbf{p}+\mathbf{q}_1+\mathbf{q}_2)\frac{1}{\sqrt{V}}\frac{i}{4}\epsilon_{abc}\frac{N_3}{2\pi},
\end{flalign}
where $N_3$ was defined in Eq. \eqref{N_3_def}.
\newpage
Putting this all together, we see that the effective action is given by,
\begin{widetext}
\begin{flalign}
    S_{\mathrm{eff}}[A]&=\frac{N_3}{8\pi}\int\frac{\mathrm{d}\omega}{2\pi}\sum_{\mathbf{p}}A^a_{\mu}(-\mathbf{p},-\omega)\epsilon^{\mu\nu\lambda}(-p_{\nu})A^a_{\lambda}(\mathbf{p},\omega)\nonumber\\
    &-\frac{1}{3\sqrt{V}}\frac{iN_3}{8\pi}\epsilon_{abc}\epsilon^{\mu\nu\lambda}\sum_{\mathbf{p},\mathbf{q}_1,\mathbf{q}_2}\int\frac{\mathrm{d}\omega\mathrm{d}\omega_1\mathrm{d}\omega_2}{(2\pi)^2}\delta(\mathbf{p}+\mathbf{q}_1+\mathbf{q}_2)\delta(\omega-\omega_1-\omega_2)A^a_{\mu}(-\mathbf{p},-\omega)A_{\nu}^b(-\mathbf{q}_1,\omega_1)A_{\lambda}^c(-\mathbf{q}_2,\omega_2)\nonumber\\
    &=\frac{N_3}{8\pi}\int\frac{\mathrm{d}\omega}{2\pi}\sum_{\mathbf{p}}A^a_{\mu}(-\mathbf{p},-\omega)\epsilon^{\mu\nu\lambda}(-p_{\nu})A^a_{\lambda}(\mathbf{p},\omega)\nonumber\\
    &-\frac{1}{3\sqrt{V}}\frac{iN_3}{8\pi}\epsilon_{abc}\epsilon^{\mu\nu\lambda}\sum_{\mathbf{p},\mathbf{q}_1,\mathbf{q}_2}\int\frac{\mathrm{d}\omega\mathrm{d}\omega_1\mathrm{d}\omega_2}{(2\pi)^3}(2\pi)\delta(\omega-\omega_1-\omega_2)\delta(\mathbf{p}-\mathbf{q}_1-\mathbf{q}_2)A^a_{\mu}(-\mathbf{p},-\omega)A_{\nu}^b(\mathbf{q}_1,\omega_1)A_{\lambda}^c(\mathbf{q}_2,\omega_2)\nonumber\\
    &=\frac{iN_3}{4\pi}\int\mathrm{Tr}_s\left(A\wedge \mathrm{d}A+\frac{2i}{3}A\wedge A\wedge A\right),
\end{flalign}
\end{widetext}
where $\mathrm{Tr}_s$ denotes trace over the spin Pauli matrices and $N_3$ is given by,
\begin{flalign}
    N_3=-\frac{2\pi}{\beta V}\frac{\epsilon^{\mu\nu\lambda}}{6}\sum_{\mathbf{Q},i\nu_F}\mathrm{Tr}\Big([G^{-1}\partial_{\mu}G][G^{-1}\partial_{\nu}G][G^{-1}\partial_{\lambda}G]\Big)
\end{flalign}
Note that, in the last line, we used the inverse Fourier transform of $A$, given by,
\begin{flalign}
    A(\mathbf{p},\omega)=\frac{1}{\sqrt{V}}\int_{-\infty}^{\infty}\mathrm{d}t\sum_{\mathbf{x}}e^{i(\omega t-\mathbf{p}\cdot\mathbf{x})}A(\mathbf{x},t).
\end{flalign}
This completes our proof.

\section{Spectral Representation}\label{sec:spectral}
We now introduce a spectral density representation allows us to extend our Lehmann representation proof to systems with continuous spectra (see Appendix \ref{Spectral density representation of Response and Matsubara Functions} for details).
Starting with the solution to the equation of motion for the density matrix,
\begin{equation}
    \rho(t)=U(t)\rho_0U^{\dagger}(t),
\end{equation}
with $U(t)=\mathbb{T}\exp\Big(-i\int_{-\infty}^t\mathrm{d}t'H(t')\Big)$, we can expand $\mathrm{tr}(A\rho_n(t))$ using the Dyson series for $U(t)$ to find 
that the \emph{unsymmetrized} causal response function is given by
\begin{flalign}
    &\tilde{\chi}_{AB_{I_1}\cdots B_{I_n}}^{(n)}(\{t-t_i\}_{i=1}^n) \nonumber\\&= (-i)^n\prod_{i=1}^n\Theta(t-t_i)\langle{[A(t),\prod_{j=1}^nB_{I_j}(t_j)]_n}\rangle_0e^{\epsilon(\sum_i t_i-t)},
\end{flalign}
where we have defined,
\begin{flalign}
    &[A(t),\prod_{j=1}^nB_{I_j}(t_j)]_n\nonumber\\&=\underbrace{\left[\left[\cdots\left[\left[A(t),B_{I_1}(t_1)\right],B_{I_2}(t_2)\right],\dots\right],B_{I_n}(t_n)\right]}_{\text{$n$ times}},
\end{flalign}
$\Theta(v)$ is the step function, and $\langle \rangle_0$ is with respect to $\rho_0$.
This generalizes the Kubo formula to arbitrary orders in perturbation theory.
The physical response function is given by symmetrizing $\tilde{\chi}^{(n)}$ over all permutations of $(I_i,t-t_i)$.
In frequency space, causality implies that $\tilde{\chi}^{(n)}_{AB_{I_1}\cdots B_{I_n}}(\{\omega_i\}_{i=1}^n)$ is an analytic function in the upper half $\omega$ planes.
This allows us to express $\tilde{\chi}^{(n)}_{AB_{I_1}\cdots B_{I_n}}(\{\omega_i\}_{i=1}^n)$ as an integral over a spectral density. 
Using the Fourier transform of the step function $\Theta(v)$, we find
\begin{flalign} 
\label{spectral_density_representation}
&\tilde{\chi}_{AB_{I_{1}}\cdots B_{I_{n}}}^{(n)}(\{\omega_{i}\}_{i=1}^n) \nonumber\\&= \left(\frac{-1}{\pi}\right)^n\int\left(\prod_{i=1}^n \frac{\mathrm{d}\alpha_i}{\alpha_i -\sum_{j=i}^n\omega_j^+}\right)\eta^{(n)}_{AB_{I_{1}}\cdots B_{I_{n}}} (\{\alpha_i\}_{i=1}^n),
\end{flalign}
where the spectral density $\eta^{(n)}$ is
\begin{flalign}
&\eta^{(n)}_{AB_{I_{1}}\cdots B_{I_{n}}}(\{\alpha_i\}_{i=1}^n)\nonumber\\&=\frac{1}{2^n}\int \prod_{i=1}^n\left(\mathrm{d}v_ie^{iv_i\alpha_i}\right)\langle[A(0),\prod_{j=1}^nB_{I_j}(-\sum_{k=1}^jv_k)]_n\rangle_0.
\end{flalign}
The full response function is given by symmetrizing Eq.~\eqref{spectral_density_representation}.
Eq.~\eqref{spectral_density_representation} generalizes the formula for the second order spectral density obtained in Ref.~\cite{bradlyn2024spectral}, and is a key result of this paper.

To see that the spectral density is a sum over delta functions, note that the right hand side of Eq.~\eqref{spectral_density_representation} is a Fourier transform of a nested commutator.
Inserting a complete set of states between each term in the nested commutator reveals that $v_i$ only appears in the form $\exp(i\epsilon_{i_1i_2}v_k)$,
the Fourier transform of which is $\delta(\alpha_k-\epsilon_{i_1i_2})$.
Therefore, the spectral density is a sum of delta functions, and Eq.~\eqref{spectral_density_representation} is equivalent to the Lehmann representation for systems with discrete spectrum.
Furthermore, the left hand side of Eq.~\eqref{spectral_density_representation} is a meromorphic function of the $\omega_i$, that is analytic in each upper half plane; the location of the poles are independent of the spectral density, allowing us to analytically continue to the complex $\omega_i$ planes independently of the spectral density.
Symmetrizing over frequencies, {we analytically continue the susceptibility to the upper half plane}
 \begin{flalign}\label{eq:Xdef_main}
 &X_{AB_{I_1}\cdots B_{I_n}}^{(n)}(\{z_i\}_{i=1}^n) \nonumber\\=& \frac{(-1)^n}{\pi^n n!}\sum_\sigma\int\prod_{i=1}^n \frac{\mathrm{d}\alpha_i}{\alpha_i -\sum_{j=i}^nz_{\sigma(j)}}\eta^{(n)}_{AB_{I_{\sigma(1)}}\cdots B_{I_{\sigma(n)}}} (\{\alpha_i\}).
 \end{flalign} 
$X_{AB}$ satisfies a recursion relation generalizing Eqs.~\eqref{recurrence_relation_for_causal_response} and \eqref{recurrence_relation_for_Matsubara_function},
\begin{align}\label{eq:generalrecurrence_main}
&\sum_{i=1}^{n}z_iX_{AB_{I_1}\cdots B_{I_n}}^{(n)}(\{z_i\}_{i=1}^n) \nonumber \\
&= X_{[A,H_0]B_{I_1}\cdots B_{I_n}}^{(n)}(\{z_i\}_{i=1}^n)\nonumber\\& + \frac{\left\{ X^{(n-1)}_{[A,B_{I_{1}}]B_{I_{2}}\cdots B_{I_{n}}}(\{z_{i}\}_{i=2}^n) + (I_1,z_1)\leftrightarrow (I_{i\neq 1},z_{i\neq 1})\right\}}{n},
\end{align}
which shows that $X_{AB_{I_1}\cdots B_{I_n}}^{(n)}(\{z_i\}_{i=1}^n)$ connects the causal response function to the Matsubara function.
Note that, after analytic continuation, $X_{AB}$ only contains the regular parts of the Matsubara function, the anomalous terms do not appear in $X$.

\section{$n$-th Order Response}\label{sec:nth}
We now introduce an explicit solution to the recursion relation Eq.~\eqref{rec1_main}. 
Using induction, it can be shown (see Appendix \ref{Explicit solution to the recursion relation} for details) that,
\begin{equation}
\label{explicit_solution_to_recursion_main}
    f^{(n)}_{i_1,\cdots,i_{n+1}}(\{\omega_i\}_{i=1}^n)=\frac{1}{n!}\sum_\sigma\sum_{a=1}^{n+1}\rho_a\prod_{i\neq a}\frac{1}{q_a-q_i},
\end{equation}
where the sum runs over all permutations $\sigma$ of $(1,\dots,n)$, $\rho_a=e^{-\beta\epsilon_{i_a}}$, and $q_a=\epsilon_{i_a}+\sum_{k=0}^a\omega_k^+$, with $\omega_0$ being an auxiliary variable that cancels out from all calculations.
This gives the Lehmann representation of the causal response function when inserted into Eq.~\eqref{Lehmann_representation}.
One can obtain a similar formula for $g^{(n)}_{i_1\cdots i_{n+1}}$ by replacing $\omega^+$ with $i\nu$.
Similar results for the Lehmann kernel of the Matsubara and retarded response functions were derived in Refs. \cite{kugler2021multipoint, halbinger2023spectral}.

As an application, we derive a formula for the asymptotic rate of $n$-th harmonic generation, $\omega_1=\cdots=\omega_n\rightarrow\infty$.
We see that, in this limit, the causal response function is given by,
\begin{flalign}
\label{SHG}
    \lim_{\omega\rightarrow\infty}\omega^n\chi^{(n)}_{AB}(\omega)=\frac{1}{n!}\langle[\underbrace{\cdots[[A,B],B],\cdots,B]}_{n-\mathrm{times}}\rangle_0,
\end{flalign}
which constitutes a generalization of the generalized f-sum rule to all orders in perturbation theory. 
We derive this result from Eq.~\eqref{explicit_solution_to_recursion_main} in Appendix \ref{Explicit solution to the recursion relation}.
We can also see it from Eq.~\eqref{eq:Xdef_main}.
In the limit $z_1=\cdots z_n\rightarrow\infty$, we can ignore the $\alpha_i$ in the denominator of Eq.~\eqref{eq:Xdef_main}, yielding
\begin{flalign}
 &X_{AB_{I_1}\cdots B_{I_n}}^{(n)}(z) \nonumber\\\sim& \frac{1}{(z\pi)^n}\frac{1}{n!}\int\left(\prod_{i=1}^n \frac{\mathrm{d}\alpha_i}{n!}\right)\sum_{\sigma}\eta^{(n)}_{AB_{I_{\sigma(1)}}\cdots B_{I_{\sigma(n)}}} (\{\alpha_i\}_{i=1}^n).
 \end{flalign}
 Integrating by parts using Eq.~\eqref{spectral_density_representation} gives
 \begin{flalign}
 \label{multi_perturbation_SHG}
    &\lim_{z\rightarrow\infty} z^nX_{AB_{I_1}\cdots B_{I_n}}^{(n)}(z)=\frac{1}{(n!)^2}\sum_{\sigma}\eta^{(0)}_{AB_{I_{\sigma(1)}}\cdots B_{I_{\sigma(n)}}}\nonumber\\&=\frac{1}{(n!)^2}\sum_{\sigma}\langle [\cdots[[A,B_{I_{\sigma(1)}}],B_{I_{\sigma(2)}}],\cdots,B_{I_{\sigma(n)}}]\rangle_0.
 \end{flalign}
Taking $B_{I_i}=B$ yields Eq.~\eqref{SHG}.  
Note that Eqs.~\eqref{SHG} and \eqref{multi_perturbation_SHG} are the leading terms in the asymptotic expansion of the causal response function in powers of the frequency,
\begin{flalign}
    \chi^{(n)}_{AB}(\omega)\sim\frac{1}{n!}\sum_{N=1}^{\infty}\frac{[\nu^{(N)}_{AB}]_n}{\omega^{N+n}}.
\end{flalign}
The coefficients $[\nu^{(N)}_{AB}]_n$ are nonlinear generalizations of the sum rules for linear response coefficients.  
Asymptotically expanding the integrals in Eq.~\eqref{eq:Xdef_main} shows that the $[\nu^{(N)}_{AB}]_n$ can be expressed them in terms of moments of the spectral density $\eta^{(n)}_{AB}$, yielding a family of generalized sum rules\footnote{see Ref. \cite{bradlyn2024spectral} for an expression of $[\nu^{(N)}_{AB}]_2$ in terms of the second order spectral density}.

\section{Conclusion}
We have demonstrated the connection between causal and Matsubara Green's functions to all orders in perturbation theory, generalizing the textbook result at $n=1$.
Our equation of motion based proof is 
physically transparent and gives formulas for the nonlinear response functions that we used to constrain their large frequency behavior.
We emphasize that our results were obtained without making any assumptions about the strength of interactions or the form of the underlying Hamiltonian; they can be used to study many-body effects in quantum materials.

Our results can also be used to compute the non-linear sum rule coefficients $[\nu^{(N)}_{AB}]_n$.
This will generalize the sum rules obtained in Refs.~\cite{bradlyn2024spectral}, enabling a study of density-density response and sum rules for the conductivity to arbitrary orders in perturbation theory, which are related to quantum geometry of occupied electron bands in non-interacting systems~\cite{souza2000polarization, torma2022superconductivity, onishi2025quantum, alexandradinata2022quantization, holder2020consequences,balut2025quantum,balut2025quantuma,verma2025quantum,verma2025framework,patankar2018resonance,bouhon2023quantum}.
Using the Matsubara formalism, this will allow for a study of the effect of disorder and interactions on quantum geometric quantities.
Incorporating such effects in the causal response formalism is difficult due to the presence of nested commutators, and one needs to resort to perturbation theory on the Keldysh contour which technically more difficult.
Our work also opens a new window into the numerical study of nonlinear response functions.
Most numerical methods on finite temperature Green's functions, such as dynamical mean field theory (DMFT), or quantum Monte Carlo (QMC), use the Matsubara formalism.
However, numerical analytic continuation of the finite temperature, imaginary time Green's functions in frequency space to obtain the causal response is an ill-conditioned problem~\cite{gubernatis1991quantum, boehnke2011orthogonal, li2016efficient, wentzell2020high, shinaoka2017compressing, shinaoka2018overcomplete, shinaoka2020sparse}. Methods such as continued fraction Pad\'e approximation and maximum entropy lead to negative spectral functions and fail to resolve spectral features at high frequencies, respectively. 
We expect that the Nevanlinna analytical continuation techniques of Refs. \cite{fei2021nevanlinna, nogaki2023nevanlinna} to higher point Matsubara Green's functions, using our results, should yield causal responses with positive and normalized spectral functions. 
We leave this for future work.

\begin{acknowledgments}
The authors thank J.~E.~Moore, P.~W.~Philips, and S.~Sondhi for helpful discussions. {The authors would also like to thank an anonymous referee for highlighting the subtlety of anomalous terms in the Matsubara correlators.}
This work was supported by the National Science Foundation under grant no.~DMR-2510219.
\end{acknowledgments}

\onecolumngrid
\appendix
\section{Derivation of the recursion relation}
\label{Derivation of the recursion relation}
In this section, we present the details of the recurrence relation satisfied by the Lehmann representation of the causal response and Matsubara functions. 

\subsection{Causal Response Functions}

We consider a time-independent Hamiltonian $H_0$ with time dependent perturbation
\begin{equation}
    H(t)=H_0+e^{\epsilon t}f(t)B,
\end{equation}
where the factor of $e^{\epsilon t}$ ensures that the perturbation dies off at $t\rightarrow -\infty$.
We will derive a recursion relation for the $n-$th order response functions.
From the Schrodinger equation of motion for $\rho$, we see that the $n$-th order density matrix satisfies 
\begin{equation}\label{eq:rhoEOM}
    \dot{\rho_n}(t)=-i[H(t),\rho(t)]=-i[H_0,\rho_n(t)]-ie^{\epsilon t}f(t)[B,\rho_{n-1}(t)].
\end{equation} 
From this, we deduce that
\begin{flalign}
\label{trace_EOM}
    \mathrm{tr}(A\dot\rho_n(t))&=-i\mathrm{tr}(A[H_0,\rho_n(t)])-ie^{\epsilon t}f(t)\mathrm{tr}(A[B,\rho_{n-1}(t)])\nonumber\\
    &=-i\mathrm{tr}([A,H_0]\rho_n(t))-ie^{\epsilon t}f(t)\mathrm{tr}([A,B]\rho_{n-1}(t)),
\end{flalign}
where in the second line, we used the cyclic property of the trace.
Now, we define the causal response function~\cite{kubo1957statistical,kadanoff1963hydrodynamic,jha1984nonlinear},
\begin{flalign}\label{eq:response_def}
    \mathrm{tr}(A\rho_n(t))&\equiv e^{n\epsilon t}\int\prod_{i=1}^n\mathrm{d}t_if(t_i)e^{-\epsilon(t-t_i)}\chi_{AB}^{(n)}(t-t_1,\cdots,t-t_n)\nonumber\\
    &=e^{n\epsilon t}\int\prod_{i=1}^n\mathrm{d}t_if(t_i)\chi_{AB}^{(n)}(t-t_1,\cdots,t-t_n)e^{-\epsilon\sum_{i=1}^n(t-t_i)}.
\end{flalign}
Causality implies that $\chi$ vanishes whenever any of its arguments is negative.
Using the convolution theorem,
\begin{flalign}
    &\int\mathrm{d}te^{i\omega t}\int\prod_{i=1}^n\mathrm{d}t_if(t_i)\chi^{(n)}_{AB}(t-t_i,\cdots,t-t_n)e^{-\epsilon\sum_{i=1}^n(t-t_i)}\nonumber\\&=\int\mathrm{d}te^{i\omega t}\int\prod_{i=1}^n\mathrm{d}t_i\int\prod_{i=1}^n\frac{\mathrm{d}\omega_i}{2\pi}f(\omega_i)e^{-i\omega_it_i}\chi^{(n)}_{AB}(t-t_i,\cdots,t-t_n)e^{-\epsilon(t-t_1)}\cdots e^{-\epsilon(t-t_n)}\nonumber\\&=\int\mathrm{d}t\int\prod_{i=1}^n\mathrm{d}t_i\Big\{\prod_{i=1}^n\frac{\mathrm{d}\omega_i}{2\pi}f(\omega_i)e^{i\omega_i(t-t_i)}e^{-\epsilon(t-t_i)}\Big\}\chi^{(n)}_{AB}(t-t_1,\cdots,t-t_n)e^{it(\omega-\sum_{i=1}^n\omega_i)}.
\end{flalign}
If we define $u_0=t$ and $u_i=t-t_i$, we have,
\begin{flalign}
    &\int\mathrm{d}t\int\prod_{i=1}^n\mathrm{d}t_i\Big\{\prod_{i=1}^n\frac{\mathrm{d}\omega_i}{2\pi}f(\omega_i)e^{i\omega_i(t-t_i)}e^{-\epsilon(t-t_i)}\Big\}\chi^{(n)}_{AB}(t-t_1,\cdots,t-t_n)e^{it(\omega-\sum_{i=1}^n\omega_i)}\nonumber\\&=\int\mathrm{d}u_0\int\prod_{i=1}^n\mathrm{d}u_i\prod_{i=1}^n\frac{\mathrm{d}\omega_i}{2\pi}f(\omega_i)e^{i\omega_i^+u_i}\chi^{(n)}_{AB}(u_1,\cdots,u_n)e^{iu_0(\omega-\sum_{i=1}^n\omega_i)}\nonumber\\
    &=\int\prod_{i=1}^n\frac{\mathrm{d}\omega_i}{2\pi}f(\omega_i)2\pi\delta\Big(\omega-\sum_{i=1}^n\omega_i\Big)\chi^{(n)}_{AB}(\omega_1^+,\cdots,\omega_n^+),
\end{flalign}
where we have used the notation $\omega^+=\omega+i\epsilon$. 
Since the integration measure is symmetric in $\omega_1,\cdots\omega_n$, only the symmetric part of $\chi_{AB}^{(n)}(\omega^+_1,\cdots,\omega_n^+)$ survives.
Therefore, in frequency space,
\begin{flalign}
    \mathrm{tr}(A\rho_n(t))&=e^{n\epsilon t}\int\frac{\mathrm{d}\omega}{2\pi} e^{-i\omega t}\int\prod_{i=1}^n\frac{\mathrm{d}\omega_i}{2\pi}f(\omega_i)2\pi\delta\Big(\omega-\sum_{i=1}^n\omega_i\Big)\chi^{(n)}_{AB}(\omega_1^+,\cdots,\omega_n^+)\nonumber\\&=\int{\mathrm{d}\omega}\int\Big\{\prod_{i=1}^n\frac{\mathrm{d}\omega_i}{2\pi}e^{-i\omega_i^+t}f(\omega_i)\Big\}\chi^{(n)}_{AB}(\omega_1^+,\cdots,\omega_n^+)\delta\Big(\omega-\sum_{i=1}^n\omega_i\Big),
\end{flalign}
which implies,
\begin{flalign}
    \mathrm{tr}(A\dot\rho_n(t))=e^{n\epsilon t}\int\mathrm{d}\omega e^{-i\omega t}\int\prod_{i=1}^n\frac{\mathrm{d}\omega_i}{2\pi}f(\omega_i)\Big(-i\sum_{i=1}^n\omega_i^+\Big)\chi^{(n)}_{AB}(\omega_1^+,\cdots,\omega_n^+)\delta\Big(\omega-\sum_{i=1}^n\omega_i\Big).
\end{flalign}
Similarly, we have,
\begin{flalign}
    \mathrm{tr}([A,H_0]\rho_n(t))=\int{\mathrm{d}\omega}e^{-i\omega t}\int\Big\{\prod_{i=1}^n\frac{\mathrm{d}\omega_i}{2\pi}f(\omega_i)\Big\}\chi^{(n)}_{[A,H_0]B}(\omega_1^+,\cdots,\omega_n^+)\delta\Big(\omega-\sum_{i=1}^n\omega_i\Big).
\end{flalign}
Finally, we evaluate the last term of Eq. \eqref{trace_EOM},
\begin{flalign}
\label{3rd_term_FT}
    &e^{\epsilon t}f(t)\mathrm{tr}([A,B]\rho_{n-1}(t))\nonumber\\&=e^{\epsilon t}f(t)\int\mathrm{d}\omega\int\prod_{i=2}^n\Big\{\frac{\mathrm{d}\omega_i}{2\pi}e^{-i\omega_i^+t}f(\omega_i)\Big\}\chi^{(n-1)}_{[A,B]B}(\omega_2^+,\cdots,\omega_n^+)\delta\Big(\omega-\sum_{i=2}^n\omega_i\Big)\nonumber\\&=\int\frac{\mathrm{d}\omega_1}{2\pi}e^{-i\omega_1^+t}f(\omega_1)\int\mathrm{d}\omega\int\prod_{i=2}^n\Big\{\frac{\mathrm{d}\omega_i}{2\pi}e^{-i\omega_i^+t}f(\omega_i)\Big\}\chi^{(n-1)}_{[A,B]B}(\omega_2^+,\cdots,\omega_n^+)\delta\Big(\omega-\sum_{i=2}^n\omega_i\Big)\nonumber\\&=\int\mathrm{d}\omega'\int\prod_{i=1}^n\Big\{\frac{\mathrm{d}\omega_i}{2\pi}e^{-i\omega_i^+t}f(\omega_i)\Big\}\chi^{(n-1)}_{[A,B]B}(\omega_2^+,\cdots,\omega_n^+)\delta\Big(\omega'-\sum_{i=1}^n\omega_i\Big),
\end{flalign}
where we have defined $\omega'=\omega+\omega_1$.
Note that since the integration measure is symmetric, we can explicitly symmetrize the last equation with respect to $\omega_1,\cdots,\omega_n$, to get,
\begin{flalign}
    &e^{\epsilon t}f(t)\mathrm{tr}([A,B]\rho_{n-1}(t))\nonumber\\&=\int\mathrm{d}\omega'\int\prod_{i=1}^n\Big\{\frac{\mathrm{d}\omega_i}{2\pi}e^{-i\omega_i^+t}f(\omega_i)\Big\}\frac{1}{n}\Big(\chi^{(n-1)}_{[A,B]B}(\omega_2^+,\cdots,\omega_n^+)+(\omega_1\leftrightarrow\omega_{i\neq 1})\Big)\delta\Big(\omega'-\sum_{i=1}^n\omega_i\Big).
\end{flalign}
Putting this all together, we get, from the equations of motion for the density matrix,
\begin{equation}
\label{characteristic_equation}
    \Big(\sum_{i=1}^n\omega_i^+\Big)\chi^{(n)}_{AB}(\omega^+_1,\cdots,\omega^+_n)=\chi^{(n)}_{[A,H_0]B}(\omega^+_1,\cdots,\omega^+_n)+\frac{1}{n}\Big\{\chi^{(n-1)}_{[A,B]B}(\omega^+_2,\cdots,\omega^+_n)+(\omega_1\leftrightarrow\omega_{i\neq 1})\Big\}.
\end{equation} 
Note that if $[A,B]=0$, the second term of the recursion relation Eq. \eqref{characteristic_equation} is zero.
In that case, we will first need to relate $\chi^{(n)}_{AB}$ to to $\chi^{(n)}_{[A,H_0]B}$ where $[A,H_0]$ and $B$ do not commute.
Then we use the recurrence relations for $\chi^{(n)}_{[A,H_0]B}$.

We now re-express Eq.~\eqref{characteristic_equation} in a basis of exact eigenstates of $H_0$. 
We use the Lehmann representation to write,
\begin{equation}
\label{Lehmann_representation_app}
    \chi^{(n)}_{AB}(\omega_1,\cdots,\omega_n)=\frac{1}{\mathcal{Z}}\sum_{i_1,\cdots,i_{n+1}}B^{i_1}_{i_2}\cdots B^{i_n}_{i_{n+1}}A^{i_{n+1}}_{i_1}f^{(n)}_{i_1,\cdots,i_{n+1}}(\omega_1,\cdots,\omega_n),
\end{equation}
where, $A^i_j=\langle i|A|j\rangle$, are the elements of $A$ in the eigenbasis of $H_0$.
The first term on the right hand side of the characteristic equation Eq. \eqref{characteristic_equation} becomes,
\begin{flalign}
    \chi^{(n)}_{[A,H_0]B}(\omega^+_1,\cdots,\omega^+_n)&=\frac{1}{\mathcal{Z}}\sum_{i_1,\cdots,i_{n+1}}B^{i_1}_{i_2}\cdots B^{i_n}_{i_{n+1}}[A,H_0]^{i_{n+1}}_{i_1}f^{(n)}_{i_1,\cdots,i_{n+1}}(\omega_1,\cdots,\omega_n)\nonumber\\&=\frac{1}{\mathcal{Z}}\sum_{i_1,\cdots,i_{n+1}}B^{i_1}_{i_2}\cdots B^{i_n}_{i_{n+1}}\Big((AH_0)^{i_{n+1}}_{i_1}-(H_0A)^{i_{n+1}}_{i_1}\Big)f^{(n)}_{i_1,\cdots,i_{n+1}}(\omega_1,\cdots,\omega_n)\nonumber\\&=\frac{1}{\mathcal{Z}}\sum_{i_1,\cdots,i_{n+1}}B^{i_1}_{i_2}\cdots B^{i_n}_{i_{n+1}}\Big(A^{i_{n+1}}_{i_1}\epsilon_{i_1}-A^{i_{n+1}}_{i_1}\epsilon_{i_{n+1}}\Big)f^{(n)}_{i_1,\cdots,i_{n+1}}(\omega_1,\cdots,\omega_n)\nonumber\\&=\frac{1}{\mathcal{Z}}\sum_{i_1,\cdots,i_{n+1}}B^{i_1}_{i_2}\cdots B^{i_n}_{i_{n+1}}A^{i_{n+1}}_{i_1}\epsilon_{i_1i_{n+1}}f^{(n)}_{i_1,\cdots,i_{n+1}}(\omega_1,\cdots,\omega_n)\nonumber,
\end{flalign}
where $\epsilon_i$ are the energy eigenvalues of $H_0$, and we have defined, $\epsilon_{ij}=\epsilon_i-\epsilon_j$.
For the second term on the right hand side of Eq. \eqref{characteristic_equation}, we have,
\begin{flalign}
    &\chi^{(n-1)}_{[A,B]B}(\omega_2,\cdots,\omega_n)\nonumber\\&=\frac{1}{\mathcal{Z}}\sum_{i_1,\cdots,i_{n}}B^{i_1}_{i_2}\cdots B^{i_{n-1}}_{i_{n}}[A,B]^{i_{n}}_{i_1}f^{(n-1)}_{i_1,\cdots,i_{n}}(\omega_2,\cdots,\omega_n)\nonumber\\&=\frac{1}{\mathcal{Z}}\sum_{i_1,\cdots,i_{n},i_{n+1}}B^{i_1}_{i_2}\cdots B^{i_{n-1}}_{i_{n}}\Big(A^{i_{n}}_{i_{n+1}}B^{i_{n+1}}_{i_1}-B^{i_{n}}_{i_{n+1}}A^{i_{n+1}}_{i_1}\Big)f^{(n-1)}_{i_1,\cdots,i_{n}}(\omega_2,\cdots,\omega_n)\nonumber\\
    &=\frac{1}{\mathcal{Z}}\sum_{i_1,\cdots,i_{n+1}}B^{i_1}_{i_2}\cdots B^{i_{n-1}}_{i_{n}}B^{i_{n}}_{i_{n+1}}A^{i_{n+1}}_{i_1}\Big(f^{(n-1)}_{i_2,\cdots,i_{n+1}}(\omega_2,\cdots,\omega_n)-f^{(n-1)}_{i_1,\cdots,i_{n}}(\omega_2,\cdots,\omega_n)\Big).
\end{flalign}
In the last line, we relabeled indices $i_{n+1}\rightarrow i_1$, $i_1\rightarrow i_2$, $\cdots$, $i_n\rightarrow i_{n+1}$.
Finally, exploiting the fact that Eq. \eqref{characteristic_equation} is true for arbitrary $A$ and $B$, we get,
\begin{equation}
\label{rec1}
    f^{(n)}_{i_1,\cdots,i_{n+1}}(\omega_1,\cdots,\omega_n)=\frac{1}{n(\omega_t^+-\epsilon_{i_1i_{n+1}})}\Bigg\{f^{(n-1)}_{i_2,\cdots,i_{n+1}}(\omega_2,\cdots,\omega_n)-f^{(n-1)}_{i_1,\cdots,i_{n}}(\omega_2,\cdots,\omega_n)+(\omega_1\leftrightarrow\omega_{i\neq 1})\Bigg\},
\end{equation}
where $\omega_t^+=\sum_{i=1}^n\omega_i^+$ and we have symmetrized the expression in the $\omega_i$s.

As an example, we can use the recursion relation Eq. \eqref{rec1} to evaluate $f$ for $n=2$. 
The first order expression is,
\begin{equation}
    f^{(1)}_{i_1,i_2}(\omega)=\frac{e^{-\beta\epsilon_{i_2}}-e^{-\beta\epsilon_{i_1}}}{\omega^+-\epsilon_{i_1i_2}}.
\end{equation}
Therefore, using the recursion relation,
\begin{equation}
\label{2nd_order_causal_response}
    f^{(2)}_{i_1,i_2,i_3}(\omega_1,\omega_2)=\frac{1}{2(\omega_{12}^+-\epsilon_{i_1i_3})}\Big\{\frac{e^{-\beta\epsilon_{i_3}}-e^{-\beta\epsilon_{i_2}}}{\omega_2^+-\epsilon_{i_2i_3}}-\frac{e^{-\beta\epsilon_{i_2}}-e^{-\beta\epsilon_{i_1}}}{\omega_2^+-\epsilon_{i_1i_2}}+(\omega_1\longleftrightarrow\omega_2)\Big\},
\end{equation}
which matches the Lehmann representation for the second-order response function derived in Ref. \cite{bradlyn2024spectral}.

\subsection{Matsubara Functions}

We can carry out an analogous derivation for the Lehmann representation of the Matsubara function
\begin{equation}
    \bar{\chi}^{(n)}_{AB}(\tau_1,\cdots,\tau_n)=\frac{1}{n!}\mathrm{tr}(\mathbb{T}A(0)B(-\tau_1)\cdots B(-\tau_n)\rho_0),
\end{equation}
where $B(\tau)=e^{\tau H_0}Be^{-\tau H_0}$, and $\mathbb{T}$ indicates imaginary time ordering.  
First, note that,
\begin{equation}
    \sum_{i=1}^n\frac{\partial}{\partial\tau_i}\bar{\chi}^{(n)}_{AB}(\tau_1,\cdots,\tau_n)=\frac{\mathrm{d}}{\mathrm{d}\tau}\bar{\chi}^{(n)}_{AB}(\tau+\tau_1,\cdots,\tau+\tau_n)\Bigg|_{\tau=0},
\end{equation}
where we used the chain rule.
By using time translation invariance of the correlation function, this gives us,
\begin{flalign}
\label{chain_rule_app}
    \sum_{i=1}^n\frac{\partial}{\partial\tau_i}\bar{\chi}^{(n)}_{AB}(\tau_1,\cdots,\tau_n)
    &=\frac{\mathrm{d}}{\mathrm{d}\tau}\bar{\chi}^{(n)}_{AB}(\tau+\tau_1,\cdots,\tau+\tau_n)\Bigg|_{\tau=0}\nonumber\\&=\frac{1}{n!}\frac{\mathrm{d}}{\mathrm{d}\tau}\mathrm{tr}(\mathbb{T}A(\tau)B(-\tau_1)\cdots B(-\tau_n)\rho_0)\Bigg|_{\tau=0}\nonumber\\&=\frac{1}{n!}\frac{\mathrm{d}}{\mathrm{d}\tau}\mathrm{tr}(A(\tau)\mathbb{T}B(-\tau_1)\cdots B(-\tau_n)\rho_0)\Bigg|_{\tau=0}\nonumber\\&=-\bar{\chi}^{(n)}_{[A,H_0]B}(\tau_1,\cdots,\tau_n),
\end{flalign}
where we used $\frac{\mathrm{d}}{\mathrm{d}\tau}A(\tau)=-[A,H_0]$.
If we now Fourier transform both sides of Eq. \eqref{chain_rule_app}, we get,
\begin{equation}
    \prod_{k=1}^n\int_0^{\beta}\mathrm{d}\tau_ke^{i\nu_k\tau_k}\sum_{i=1}^n\frac{\partial}{\partial\tau_i}\bar{\chi}^{(n)}_{AB}(\tau_1,\cdots,\tau_n)=-\bar{\chi}^{(n)}_{[A,H_0]B}(\nu_1,\cdots,\nu_n).
\end{equation}
In the region of integration $\mathrm{tr}(\mathbb{T}A(0)B(-\tau_1)\cdots B(-\tau_n)\rho_0)=\mathrm{tr}(A\mathbb{T}B(-\tau_1)\cdots B(-\tau_n)\rho_0)$, since all the $\tau$s are positive.
Furthermore, integrating by parts yields
\begin{flalign}
    \int_0^{\beta}\mathrm{d}\tau_ke^{i\nu_k\tau_k}\frac{\partial}{\partial\tau_k}\bar{\chi}^{(n)}_{AB}(\tau_1,\cdots,\tau_n)=&-i\nu_k\bar{\chi}^{(n)}_{AB}(\tau_1,\cdots,\nu_k,\cdots,\tau_n)\nonumber\\&+\bar{\chi}^{(n)}_{AB}(\tau_1,\cdots,\beta,\cdots,\tau_n)-\bar{\chi}^{(n)}_{AB}(\tau_1,\cdots,0,\cdots,\tau_n).
\end{flalign}
Using the time ordering symbol, we have,
\begin{flalign}
    \bar{\chi}^{(n)}_{AB}(\tau_1,\cdots,0,\cdots,\tau_n)&=\frac{1}{n!}\mathrm{tr}(A\mathbb{T}B(-\tau_1)\cdots B(0)\cdots B(-\tau_n)\rho_0)\nonumber\\&=\frac{1}{n!}\mathrm{tr}(AB\mathbb{T}B(-\tau_1)\cdots \xcancel{B(\tau_k)}\cdots B(-\tau_n)\rho_0)\nonumber\\&=\frac{1}{n}\bar{\chi}^{(n-1)}_{(AB)B}(\tau_1,\cdots,\xcancel{\tau_k},\cdots,\tau_n)
\end{flalign}
and similarly,
\begin{flalign}
    \bar{\chi}^{(n)}_{AB}(\tau_1,\cdots,\beta,\cdots,\tau_n)&=\frac{1}{n!}\mathrm{tr}(A\mathbb{T}B(-\tau_1)\cdots B(-\beta)\cdots B(-\tau_n)\rho_0)\nonumber\\&=\frac{1}{n!}\mathrm{tr}(A\mathbb{T}B(-\tau_1)\cdots B(-\tau_n)\rho_0 B\rho^{-1}_0\rho_0)\nonumber\\&=\frac{1}{n!}\mathrm{tr}(BA\mathbb{T}B(-\tau_1)\cdots B(-\tau_n)\rho_0)\nonumber\\
    &=\frac{1}{n}\bar{\chi}^{(n-1)}_{(BA)B}(\tau_1,\cdots,\xcancel{\tau_k},\cdots,\tau_n).
\end{flalign}
Therefore, we have,
\begin{equation}
    \prod_{k=1}^n\int_0^{\beta}\mathrm{d}\tau_ke^{i\nu_k\tau_k}\sum_{i=1}^n\frac{\partial}{\partial\tau_i}\bar{\chi}^{(n)}_{AB}(\tau_1,\cdots,\tau_n)=\Big(-i\sum_{i=1}^n\nu_i\Big)\bar{\chi}^{(n)}_{AB}(\nu_1,\cdots.\nu_n)-\frac{1}{n}\sum_{i=1}^n\bar{\chi}^{(n-1)}_{[A,B]B}(\nu_1,\cdots,\xcancel{\nu_i},\cdots,\nu_n)\nonumber.
\end{equation}
We can now exploit the symmetry of $\bar{\chi}^{(n-1)}$ in its frequency arguments to rewrite the sum in the last line in terms of a symmetrization over $\nu_1\leftrightarrow\nu_i$ analogously to Eq.~\eqref{characteristic_equation}.
The characteristic equation for the Matsubara Green functions then becomes,
\begin{equation}\label{eq:matsubara_characteristic}
    -i\Big(\sum_{i=1}^n\nu_i\Big)\bar{\chi}^{(n)}_{AB}(\nu_1,\cdots,\nu_n)=-\bar{\chi}^{(n)}_{[A,H_0]B}(\nu_1,\cdots,\nu_n)+\frac{1}{n}\Big\{\bar{\chi}^{(n-1)}_{[A,B]B}(\nu_2,\cdots,\nu_n)+(\nu_1\longleftrightarrow\nu_{i\neq 1})\Big\}.
\end{equation}
Now, we can use the Lehmann representation again, to write,
\begin{equation}
    \bar{\chi}^{(n)}_{AB}(\nu_1,\cdots,\nu_n)=\frac{1}{\mathcal{Z}}\sum_{i_1,\cdots,i_{n+1}}B^{i_1}_{i_2}\cdots B^{i_n}_{i_{n+1}}A^{i_{n+1}}_{i_1}g^{(n)}_{i_1,\cdots,i_{n+1}}(\nu_1,\cdots,\nu_n).
\end{equation}
Plugging this into Eq.~\eqref{eq:matsubara_characteristic}, we get the recurrence relation
\begin{equation}
\label{rec2}
    g^{(n)}_{i_1,\cdots,i_{n+1}}(\nu_1,\cdots,\nu_n)=\frac{-1}{n(i\nu_t-\epsilon_{i_1i_{n+1}})}\Bigg\{g^{(n-1)}_{i_2,\cdots,i_{n+1}}(\nu_2,\cdots,\nu_n)-g^{(n-1)}_{i_1,\cdots,i_{n}}(\nu_2,\cdots,\nu_n)+(\nu_1\longleftrightarrow\nu_{i\neq 1})\Bigg\}
\end{equation}
We will use induction to show the following result to all orders in perturbation theory,
\begin{equation}\label{eq:lehmann-equivalence}
    f^{(n)}_{i_1,\cdots,i_{n+1}}(\omega_1,\cdots,\omega_n)=(-1)^ng^{(n)}_{i_1,\cdots,i_{n+1}}(i\nu_1=\omega_1^+,\cdots,i\nu_n=\omega_n^+).
\end{equation}
Note that, we have the well known result at $n=1$ that
\begin{equation}
    f^{(1)}_{i_1,i_2}(\omega)=-g^{(1)}_{i_1,i_2}(i\nu=\omega+i\epsilon).
\end{equation}
This is because,
\begin{flalign}
    \int_0^{\beta}\mathrm{d}\tau e^{i\nu\tau}\mathrm{tr}(AB(-\tau)\rho_0)&=\frac{1}{\mathcal{Z}}\sum_{i_1,i_2}B^{i_1}_{i_2}A^{i_2}_{i_1}e^{-\beta\epsilon_{i_2}}\frac{e^{\beta\epsilon_{i_2i_1}}-1}{i\nu+\epsilon_{i_2i_1}}\nonumber\\&=-\frac{1}{\mathcal{Z}}\sum_{i_1,i_2}B^{i_1}_{i_2}A^{i_2}_{i_1}\frac{e^{-\beta\epsilon_{i_2}}-e^{-\beta\epsilon_{i_1}}}{i\nu-\epsilon_{i_1i_2}}\nonumber\\&=-\frac{1}{\mathcal{Z}}\sum_{i_1,i_2}B^{i_1}_{i_2}A^{i_2}_{i_1}f^{(1)}_{i_1,i_2}(\omega^+=i\nu)\nonumber.
\end{flalign}
Then, we assume that for the $n-1$th step, the following equality holds,
\begin{equation}
    f^{(n-1)}_{i_1,\cdots,i_{n}}(\omega_1,\cdots,\omega_{n-1})=(-1)^{(n-1)}g^{(n-1)}_{i_1,\cdots,i_{n}}(i\nu_1=\omega_1^+,\cdots,i\nu_{n-1}=\omega_{n-1}^+).
\end{equation}
Finally, if we compare the forms of the equations, \eqref{rec1} and \eqref{rec2}, we see that the response functions are equal in the $n$-th step as well. 
Thus, by induction, Eq.~\eqref{eq:lehmann-equivalence} holds.

\section{Direct calculation of the second order response function}
\label{Direct calculation of the second order response function}
Here, we show that the second order causal response and the second order Matsubara response functions are related by direct computation.
To second order in the perturbation $f(t)$, the response of an operator $A$ in frequency space is,
\begin{equation}
    \delta^2\langle A\rangle(\omega)=\int\frac{\mathrm{d}\omega_1}{2\pi}\frac{\mathrm{d}\omega_2}{2\pi}\delta(\omega-\omega_1-\omega_2)\chi^{(2)}_{AB}(\omega_1,\omega_2)f(\omega_1)f(\omega_2),
\end{equation}
where,
\begin{equation}
\label{2nd_ord_chi} 
    \chi^{(2)}_{AB}(\omega_1,\omega_2)=\frac{1}{2\mathcal{Z}}\sum_{nml}B^l_nB^n_mA^m_l\Bigg[\frac{e^{-\beta\epsilon_l}-e^{-\beta\epsilon_n}}{(\epsilon_{lm}-\omega_{12}^+)(\epsilon_{ln}-\omega_2^+)}+\frac{e^{-\beta\epsilon_m}-e^{-\beta\epsilon_n}}{(\epsilon_{lm}-\omega_{12}^+)(\epsilon_{nm}-\omega_2^+)}\Bigg]+(\omega_1\longleftrightarrow\omega_2).
\end{equation}
This is the same as Eq. \eqref{2nd_order_causal_response} if we replace $l\rightarrow i_1$, $n\rightarrow i_2$ and $m\rightarrow i_3$.
We will now show that this is equivalent to the Matsubara function
\begin{equation}\label{eq:second_order_matsubara}
    \bar{\chi}^{(2)}_{AB}(\nu_1,\nu_2)=\frac{1}{2}\int_0^{\beta}\mathrm{d}\tau_1e^{i\nu_1\tau_1}\int_0^{\beta}\mathrm{d}\tau_2e^{i\nu_2\tau_2}\langle\mathbb{T}A(\tau_1+\tau_2)B(\tau_1)B(\tau_2)\rangle\Bigg|_{i\nu_i=\omega_i}
\end{equation}
after analytic continuation. 

First, consider $\tau_2>\tau_1$. 
The time ordering symbol lets us write Eq.~\eqref{eq:second_order_matsubara} as
\begin{flalign*}
    &\int_0^{\beta}\mathrm{d}\tau_2e^{i\nu_2\tau_2}\int_0^{\tau_2}\mathrm{d}\tau_1e^{i\nu_1\tau_1}\langle A(\tau_1+\tau_2)B(\tau_2)B(\tau_1)\rangle\\&
    =\frac{1}{\mathcal{Z}}\sum_{nml}A^l_nB^n_mB^m_le^{-\beta\epsilon_l}\int_0^{\beta}\mathrm{d}\tau_2e^{(i\nu_2+\epsilon_{lm})\tau_2}\int_0^{\tau_2}\mathrm{d}\tau_1e^{(i\nu_1+\epsilon_{mn})\tau_1}\\&
    =\frac{1}{\mathcal{Z}}\sum_{nml}A^l_nB^n_mB^m_le^{-\beta\epsilon_l}\int_0^{\beta}\mathrm{d}\tau_2\frac{e^{(i\nu_{12}+\epsilon_{ln})\tau_2}-e^{(i\nu_2+\epsilon_{lm})\tau_2}}{i\nu_1+\epsilon_{mn}}\\&
    =\frac{1}{\mathcal{Z}}\sum_{nml}A^l_nB^n_mB^m_le^{-\beta\epsilon_l}\Bigg[\frac{e^{\beta\epsilon_{ln}}-1}{(i\nu_{12}+\epsilon_{ln})(i\nu_1+\epsilon_{mn})}-\frac{e^{\beta\epsilon_{lm}}-1}{(i\nu_{2}+\epsilon_{lm})(i\nu_1+\epsilon_{mn})}\Bigg]\\&
    =\frac{1}{\mathcal{Z}}\sum_{nml}B^l_nB^n_mA^m_l\Bigg[\frac{e^{-\beta\epsilon_{l}}-e^{-\beta\epsilon_m}}{(\epsilon_{lm}-i\nu_{12})(\epsilon_{ln}-i\nu_1)}-\frac{e^{-\beta\epsilon_{n}}-e^{-\beta\epsilon_m}}{(\epsilon_{nm}-i\nu_{2})(\epsilon_{ln}-i\nu_1)}\Bigg]\\
    &=\frac{1}{\mathcal{Z}}\sum_{nml}\frac{B^l_nB^n_mA^m_l}{\epsilon_{ln}-i\nu_1}\Bigg[\frac{e^{-\beta\epsilon_{l}}-e^{-\beta\epsilon_n}+e^{-\beta\epsilon_n}-e^{-\beta\epsilon_m}}{(\epsilon_{lm}-i\nu_{12})}-\frac{e^{-\beta\epsilon_{n}}-e^{-\beta\epsilon_m}}{(\epsilon_{nm}-i\nu_{2})}\Bigg]\\
    &=\frac{1}{\mathcal{Z}}\sum_{nml}\frac{B^l_nB^n_mA^m_l}{\epsilon_{ln}-i\nu_1}\Bigg[\frac{e^{-\beta\epsilon_l}-e^{-\beta\epsilon_n}}{(\epsilon_{lm}-i\nu_{12})}+(e^{-\beta\epsilon_n}-e^{-\beta\epsilon_m})\Bigg\{\frac{1}{\epsilon_{lm}-i\nu_{12}}-\frac{1}{\epsilon_{nm}-i\nu_2}\Bigg\}\Bigg]\\&
    =\frac{1}{\mathcal{Z}}\sum_{nml}B^l_nB^n_mA^m_l\Bigg[\frac{e^{-\beta\epsilon_l}-e^{-\beta\epsilon_n}}{(\epsilon_{lm}-i\nu_{12})(\epsilon_{ln}-i\nu_{1})}+\frac{e^{-\beta\epsilon_m}-e^{-\beta\epsilon_n}}{(\epsilon_{lm}-i\nu_{12})(\epsilon_{nm}-i\nu_2)}\Bigg],
\end{flalign*}
where we have used that $e^{i\nu_i\beta}=1$, since $\nu_i$s are Matsubara frequencies.
We have also defined $\nu_{12}=\nu_1+\nu_2$.
For $\tau_1>\tau_2$, we get the same answer except with $\nu_1\longleftrightarrow\nu_2$. 
Therefore, when we add the two results, we get by comparing with Eq.~\eqref{2nd_ord_chi}
\begin{equation}
\bar{\chi}^{(2)}_{AB}(\nu_1,\nu_2) = \chi^{(2)}_{AB}(\omega_1^+\rightarrow i\nu_1,\omega_2^+\rightarrow i\nu_2)
\end{equation}
as desired.

\section{Explicit solution to the recursion relation}
\label{Explicit solution to the recursion relation}
Now, we will derive an explicit formula for $f^{(n)}_{i_1,\cdots,i_{n+1}}(\omega_1,\cdots,\omega_n)$ by solving the recurrence relation Eq.~\eqref{characteristic_equation}.
First, we note that the symmetrization in Eq. \eqref{characteristic_equation} is unnecessary for the recurrence relation, since it was derived from Eq.~\eqref{trace_EOM} which is automatically symmetrized
In other words, we can use,
\begin{equation}
    f^{(n)}_{i_1,\cdots,i_{n+1}}(\omega_1,\cdots,\omega_n)=\frac{1}{\omega_t+\epsilon_{i_{n+1}}-\epsilon_{i_1}}\Big\{f^{(n-1)}_{i_2,\cdots,i_{n+1}}(\omega_2,\cdots,\omega_n)-f^{(n-1)}_{i_1,\cdots,i_{n}}(\omega_1,\cdots,\omega_{n-1})\Big\},
\end{equation}
where we symmetrize over all permutations of $\omega$'s at the end of the calculation.
To verify this, we see from Eq. \eqref{characteristic_equation} that if we do not explicitly symmetrize, we get,
\begin{equation}
    \omega_t^+\chi_{AB}^{(n)}(\omega_1,\cdots,\omega_n)=\chi_{[A,H_0]B}^{(n)}(\omega_1,\cdots,\omega_n)+\chi^{(n-1)}_{[A,B]B}(\omega_2,\cdots,\omega_n).
\end{equation}
We can exploit the relabeling of integration variables in Eq. \eqref{3rd_term_FT} to get,
\begin{equation}
    \omega_t^+\chi_{AB}^{(n)}(\omega_1,\cdots,\omega_n)=\chi_{[A,H_0]B}^{(n)}(\omega_1,\cdots,\omega_n)+\chi^{(n-1)}_{[A,B]B}(\omega_1,\cdots,\omega_{n-1}).
\end{equation}
Using $\chi^{(n)}_{AB}(\omega_1,\cdots,\omega_n)=\frac{1}{\mathcal{Z}}\sum_{i_1,\cdots,i_{n+1}}B^{i_1}_{i_2}\cdots B^{i_n}_{i_{n+1}}A^{i_{n+1}}_{i_1}f^{(n)}_{i_1,\cdots,i_{n+1}}(\omega_1,\cdots,\omega_n)$, which follows from Eq. \eqref{Lehmann_representation_app}, 
\begin{equation}
\label{unsymmetrized_recursion_relation}
\Big(\omega_t^+-(\epsilon_{i_1}-\epsilon_{i_{n+1}})\Big)f^{(n)}_{i_1,\cdots,i_{n+1}}(\omega_1,\cdots,\omega_n)=f^{(n-1)}_{i_2,\cdots,i_{n+1}}(\omega_1,\cdots,\omega_{n-1})-f^{(n-1)}_{i_1,\cdots,i_{n}}(\omega_1,\cdots,\omega_{n-1}).
\end{equation}
In the second term on the right hand side of Eq.~\eqref{unsymmetrized_recursion_relation}, we exploited the relabeling
\begin{flalign}
    &\chi^{(n-1)}_{[A,B]B}(\omega_1,\cdots,\omega_{n-1})=\frac{1}{\mathcal{Z}}\sum_{i_1,\cdots,i_{n}}B^{i_1}_{i_2}\cdots B^{i_n}_{i_{n}}[A,B]^{i_{n}}_{i_1}f^{(n-1)}_{i_1,\cdots,i_n}(\omega_1,\cdots,\omega_{n-1})\nonumber\\
    &=\frac{1}{\mathcal{Z}}\sum_{i_1,\cdots,i_{n+1}}\Big(B^{i_1}_{i_2}\cdots B^{i_n}_{i_{n}}A^{i_n}_{i_{n+1}}B^{i_{n+1}}_{i_1}
    -B^{i_1}_{i_2}\cdots B^{i_n}_{i_{n}}B^{i_n}_{i_{n+1}}A^{i_{n+1}}_{i_1}\Big)f^{(n-1)}_{i_1,\cdots,i_n}(\omega_1,\cdots,\omega_{n-1})\nonumber\\&=\frac{1}{\mathcal{Z}}\sum_{i_1,\cdots,i_{n+1}}B^{i_1}_{i_2}\cdots B^{i_n}_{i_{n}}B^{i_n}_{i_{n+1}}A^{i_{n+1}}_{i_1}\Big(f^{(n-1)}_{i_2,\cdots,i_{n+1}}(\omega_1,\cdots,\omega_{n-1})-f^{(n-1)}_{i_1,\cdots,i_n}(\omega_1,\cdots,\omega_{n-1})\Big).
\end{flalign}
The key point now is that we are allowed to change the arguments of $f^{(n-1)}_{i_1,\cdots,i_n}(\omega_1,\cdots,\omega_{n-1})$ in the last term of Eq.~\eqref{unsymmetrized_recursion_relation} because the defining equation, Eq. \eqref{characteristic_equation} is explicitly symmetric. 
In particular, writing
\begin{equation}
    \int\prod_{i=1}^n\mathrm{d}\omega_i\delta(\omega_t-\omega)f(\omega_i)[\omega^+_t\chi^{(n)}_{AB}(\omega_1,\cdots,\omega_n)-\chi^{(n)}_{[A,H_0]B}(\omega_1,\cdots,\omega_n)-\chi^{(n-1)}_{[A,B]B}(\omega_2,\cdots,\omega_n)]=0
\end{equation}
allows us to write the recurrence relation Eq. \eqref{unsymmetrized_recursion_relation} as
\begin{equation}\label{eq:solvable_recurrence}
    f^{(n)}_{i_1,\cdots,i_{n+1}}(\omega_1,\cdots,\omega_n)=\frac{1}{\omega^+_t+\epsilon_{i_{n+1}}-\epsilon_{i_1}}\Big(f^{(n-1)}_{i_2,\cdots,i_{n+1}}(\omega_2,\cdots,\omega_{n})-f^{(n-1)}_{i_1,\cdots,i_{n}}(\omega_1,\cdots,\omega_{n-1})\Big),
\end{equation}
where we symmetrize everything at the end.

To solve Eq.~\eqref{eq:solvable_recurrence}, we define $q_a=\epsilon_{i_a}+\sum_{k=0}^a\omega_k^+$, where we introduce an auxiliary $\omega_0^+$ that cancels out of all results.
This lets us rewrite Eq.~\eqref{eq:solvable_recurrence} as
\begin{equation}
    f^{(n)}_{i_1,\cdots,i_{n+1}}(\omega_1,\cdots,\omega_n)=\frac{1}{q_{n+1}-q_1}\Big(f^{(n-1)}_{i_2,\cdots,i_{n+1}}(\omega_2,\cdots,\omega_{n})-f^{(n-1)}_{i_1,\cdots,i_{n}}(\omega_1,\cdots,\omega_{n-1})\Big).
\end{equation}
We will show that
\begin{equation}
\label{explicit_solution_to_recursion_appendix}
    f^{(n)}_{i_1,\cdots,i_{n+1}}(\omega_1,\cdots,\omega_n)=S_{i_1,\cdots,i_{n+1}}^{(n)}(q_1^{\sigma},\cdots,q^{\sigma}_{n+1})\equiv \sum_{a=1}^{n+1}\rho_a\prod_{i\neq a}\frac{1}{q_a-q_i},
\end{equation}
where, $\rho_a=\exp(-\beta\epsilon_{i_a})$, which explicitly solves the recursion relation. Similar expressions were found in Refs.~\cite{kugler2021multipoint,ge2024analytic}. 

We now prove that Eq.~\eqref{explicit_solution_to_recursion_appendix} solves the recurrence relation using induction.
To begin, we check that for $n=1$, 
\begin{flalign}
    S^{(1)}_{i_1,i_2}(q_1,q_2)=\frac{\rho_2-\rho_1}{q_2-q_1}=\frac{\rho_1}{q_1-q_2}+\frac{\rho_2}{q_2-q_1}=f^{(1)}_{i_1,i_2}(\omega_1).
\end{flalign}
Therefore, Eq.~\eqref{explicit_solution_to_recursion_appendix} holds for $n=1$. 
Now, assume that q.~\eqref{explicit_solution_to_recursion_appendix} is true for $n-1$,
\begin{equation}
    f^{(n-1)}_{i_1,\cdots,i_{n}}(\omega_1,\cdots,\omega_{n-1})=\sum_{a=1}^{n}\rho_a\prod_{i\neq a}\frac{1}{q_a-q_i}=S^{(n-1)}_{i_1,\cdots,i_{n}}(q_1,\cdots,q_{n}).
\end{equation}
Inserting this into Eq.~\eqref{eq:solvable_recurrence}, we have
\begin{flalign}
    &f^{(n)}_{i_1,\cdots,i_{n+1}}(\omega_1,\cdots,\omega_n)=\frac{1}{q_{n+1}-q_1}\Big(f^{(n-1)}_{i_2,\cdots,i_{n+1}}(\omega_2,\cdots,\omega_{n})-f^{(n-1)}_{i_1,\cdots,i_{n}}(\omega_1,\cdots,\omega_{n-1})\Big)\nonumber\\
    &=\frac{1}{q_{n+1}-q_1}\Bigg(\sum_{a=2}^{n+1}\rho_{a}\prod_{\substack{i\neq a\\i=2}}^{n+1}\frac{1}{q_{a}-q_{i}}-\sum_{a=1}^{n}\rho_a\prod_{\substack{i\neq a\\i=1}}^n\frac{1}{q_a-q_i}\Bigg)\nonumber\\
    &=\frac{1}{q_{n+1}-q_1}\Bigg[-\rho_1\prod_{\substack{i=2}}^n\frac{1}{q_1-q_i}+\rho_{n+1}\prod_{i=1}^{n}\frac{1}{q_{n+1}-q_i}+\sum_{a=2}^n\rho_a\Big(\prod_{\substack{i\neq a\\i=2}}^{n+1}\frac{1}{q_a-q_i}-\prod_{\substack{i\neq a\\i=1}}^{n}\frac{1}{q_a-q_i}\Big)\Bigg]\nonumber\\
    &=\rho_1\prod_{i\neq 1}\frac{1}{q_1-q_i}+\rho_{n+1}\prod_{i\neq n+1}\frac{1}{q_{n+1}-q_i}+\frac{q_{n+1}-q_1}{q_{n+1}-q_1}\sum_{a=2}^n\rho_a\prod_{i\neq a}\frac{1}{q_a-q_i}\nonumber\\
    &=\sum_{a=1}^{n+1}\rho_a\prod_{i\neq a}\frac{1}{q_a-q_i}\equiv S^{(n)}_{i_1,\cdots,i_{n+1}}(q_1,\cdots,q_{n+1}).
\end{flalign}
Thus, by induction Eq.~\eqref{explicit_solution_to_recursion_appendix} solves Eq.~\eqref{eq:solvable_recurrence} for all $n$.
Upon explicit symmetrization, this means that,
\begin{equation}\label{eq:chi-lehmann-rep}
    \chi^{(n)}_{AB}(\omega_1,\cdots,\omega_n)=\frac{1}{\mathcal{Z}}\frac{1}{n!}\sum_{i_1,\cdots,i_{n+1}}B^{i_1}_{i_2}\cdots B^{i_n}_{i_{n+1}}A^{i_{n+1}}_{i_1}\sum_{\sigma}S_{i_1,\cdots,i_{n+1}}^{(n)}(q_1^{\sigma},\cdots,q^{\sigma}_{n+1}),
\end{equation}
where $\sigma$ is a permutation of $(1,\cdots,n)$ and $q^{\sigma}_a=\epsilon_{i_a}+\sum_{k=0}^{a-1}\omega_{\sigma(k)}$.

\subsection{Large frequency asymptotics}

We apply can apply the representation Eq.~\eqref{eq:chi-lehmann-rep} to derive the universal asymptotics for $n$-th harmonic generation. 
In particular, we will show that
\begin{equation}
    \lim_{\omega\rightarrow\infty}\omega^n\chi^{(n)}_{AB}(\omega,\cdots,\omega)=\frac{1}{n!}\langle[\underbrace{\cdots[[A,B],B],\cdots,B]}_{n-\mathrm{times}}\rangle_0.
\end{equation}
To see this, first note that when all the $\omega$s are equal in Eq.~\eqref{eq:chi-lehmann-rep}, we have $q_a=\epsilon_{i_a}+(a+1)\omega^+$.
In the $\omega\rightarrow\infty$ limit, the leading order Taylor expansion of the denominators of Eq.~\eqref{explicit_solution_to_recursion_appendix} is obtained by neglecting the dependence of $q_a$ on $\epsilon_{i_a}$. 
Then, from Eq. \eqref{explicit_solution_to_recursion_appendix} we find,
\begin{equation}
    f^{(n)}_{i_1,\cdots,i_{n+1}}(\omega,\cdots,\omega)\sim \sum_{a=1}^{n+1}\rho_a\prod_{i\neq a}\frac{1}{(a-i)\omega}=\frac{1}{(\omega)^n}\sum_{a=1}^{n+1}\rho_a\prod_{i\neq a}\frac{1}{(a-i)}.
\end{equation}
Therefore, we have,
\begin{flalign}
 \label{asymptotic_form}   \lim_{\omega\rightarrow\infty}\omega^nf^{(n)}_{i_1,\cdots,i_{n+1}}(\omega,\cdots,\omega)=\sum_{a=1}^{n+1}\rho_a\prod_{i\neq a}\frac{1}{(a-i)}&=\sum_{a=1}^{n+1}\rho_a\Bigg(\frac{1}{(a-1)\times\cdots\times 1}\Bigg)\Bigg(\frac{1}{-1\times\cdots\times(a-n-1)}\Bigg)\nonumber\\&=\frac{1}{n!}\sum_{a=1}^{n+1}\rho_a(-1)^{n-a+1}\Bigg(\frac{1}{(a-1)\times\cdots\times 1}\Bigg)\Bigg(\frac{n!}{1\times\cdots\times(n+1-a)}\Bigg)\nonumber\\&=\frac{1}{n!}\sum_{a=1}^{n+1}(-1)^{n-a+1}{n\choose a-1}\rho_a\nonumber\\&=\frac{1}{n!}\sum_{m=0}^{n}(-1)^{n-m}{n\choose m}\rho_{m+1}.
\end{flalign}
Therefore, we have,
\begin{equation}\label{eq:asymptotic_intermediate}
    \lim_{\omega\rightarrow\infty}\omega^n\sum_{\sigma}f^{(n)}_{i_1,\cdots,i_{n+1}}(\omega,\cdots,\omega)=\sum_{m=0}^{n}(-1)^{n-m}{n\choose m}\rho_{m+1},
\end{equation}
since Eq. \eqref{asymptotic_form} is already symmetrized, and we get an $n!$ from the sum over the $n!$ permutations.
We will also make use of the nested commutator identity~\cite{volkin1968iterated}
\begin{equation}
\label{commutator_relation}
    \underbrace{[\cdots[[A,B],B],\cdots,B]}_{n-\mathrm{times}}=\sum_{m=0}^n{n\choose m}(-1)^{n-m}B^{n-m}AB^m,
\end{equation}
which we prove using induction.
For $n=1$, the relation holds since,
\begin{equation}
    [A,B]=-BA+AB.
\end{equation}
Next, let Eq. \eqref{commutator_relation} be true for some $n$.
Then, we have,
\begin{flalign*}
    &\underbrace{[\cdots[[A,B],B],\cdots,B]}_{(n+1)-\mathrm{times}}\nonumber\\&=\sum_{m=0}^n{n\choose m}(-1)^{n-m}[B^{n-m}AB^m,B]\nonumber\\
    &=\sum_{m=0}^n{n\choose m}(-1)^{n-m}\Big(B^{n-m}AB^{m+1}-B^{n-m+1}AB^m\Big)\nonumber\\
    &=\sum_{m=1}^{n+1}(-1)^{n+1-m}{n\choose m-1}B^{n+1-m}AB^m+\sum_{m=0}^n(-1)^{n+1-m}{n\choose m}B^{n+1-m}AB^m\nonumber\\
    &=\sum_{m=1}^n\Bigg[{n\choose m-1}+{n\choose m}\Bigg](-1)^{n+1-m}B^{n+1-m}AB^m+AB^{n+1}+(-1)^{n+1}B^{n+1}A\nonumber\\
    &=\sum_{m=1}^n{n+1\choose m}(-1)^{n+1-m}B^{n+1-m}AB^m\nonumber\\&+(-1)^{n+1-(n+1)}B^{n+1-(n+1)}AB^{n+1}+(-1)^{n+1-0}B^{n+1-0}AB^0
    =\sum_{m=0}^{n+1}{n+1\choose m}(-1)^{n+1-m}B^{n+1-m}AB^m.
\end{flalign*}
Therefore, the relation holds for all $n$, by induction.
Thus, combining Eqs.~\eqref{commutator_relation} with Eq.~\eqref{eq:asymptotic_intermediate} we arrive at the desired result,
\begin{flalign}
    &\lim_{\omega\rightarrow\infty}\omega^n\chi^{(n)}_{AB}(\omega,\cdots,\omega)\nonumber\\&=\frac{1}{n!}\frac{1}{\mathcal{Z}}\sum_{i_1,\cdots,i_{n+1}}B^{i_1}_{i_2}\cdots B^{i_n}_{i_{n+1}}A^{i_{n+1}}_{i_1}n!\lim_{\omega\rightarrow\infty}\omega^n\sum_{\sigma}f^{(n)}_{i_1,\cdots,i_{n+1}}(\omega,\cdots,\omega)\nonumber\\
    &=\frac{1}{n!}\frac{1}{\mathcal{Z}}\sum_{i_1,\cdots,i_{n+1}}B^{i_1}_{i_2}\cdots B^{i_n}_{i_{n+1}}A^{i_{n+1}}_{i_1}\sum_{m=0}^n{n\choose m}(-1)^{n-m}\rho_{{m+1}}\nonumber\\
    &=\frac{1}{n!}\frac{1}{\mathcal{Z}}\sum_{m=0}^n{n\choose m}(-1)^{n-m}\sum_{i_1,\cdots,i_{n+1}}B^{i_1}_{i_2}\cdots B^{i_m}_{i_{m+1}}\rho_{{m+1}}B^{i_{m+1}}_{i_{m+2}}\cdots B^{i_n}_{i_{n+1}}A^{i_{n+1}}_{i_1}\nonumber\\
    &=\frac{1}{n!}\sum_{m=0}^n{n\choose m}(-1)^{n-m}\frac{1}{\mathcal{Z}}\mathrm{tr}\big(B^m\rho B^{n-m}A\big),
\end{flalign}
where $\rho=e^{-\beta H}$. 
In the third line, we used that 
Finally using the cyclic property of the trace to rewrite, $\mathrm{tr}\big(B^m\rho B^{n-m}A)=\mathrm{tr}\big(B^{n-m}AB^m\rho)$, we get,
\begin{equation}
    \lim_{\omega\rightarrow\infty}\omega^n\chi^{(n)}_{AB}(\omega,\cdots,\omega)=\frac{1}{n!}\langle[\cdots[[A,B],B],\cdots,B]\rangle_0.
\end{equation}
\section{Multiple perturbations}
\label{Multiple perturbations}
In this section we generalize our result to time dependent Hamiltonians of the form,
\begin{equation}
    H(t)=H_0+e^{\epsilon t}\sum_If_I(t)B_I,
\end{equation}
where $I$ indexes a set of perturbing fields (for example, Fourier components of a spatially-dependent local operator, or components of a tensor operator). 
The equation of motion for $\rho_n(t)$ is now given by
\begin{equation}\label{eq:rhoEOM_multiple}
    \dot{\rho_n}(t)=-i[H_0,\rho_n(t)]-ie^{\epsilon t}\sum_If_I(t)[B_I,\rho_{n-1}(t)].
\end{equation}
Analogous to Eq.~\eqref{eq:response_def}, we define the causal response function as
\begin{flalign}\label{eq:responsedef_multiple_t}
    \mathrm{tr}(A\rho_n(t))&\equiv e^{n\epsilon t}\sum_{\mathclap{\;\;\;\{I_i\}_{i=1}^n}}\int\prod_{i=1}^n\Big\{\mathrm{d}t_if_{I_i}(t_i)e^{-\epsilon(t-t_i)}\Big\}\chi_{AB_{I_1}\cdots B_{I_n}}^{(n)}(t-t_1,\cdots,t-t_n)\nonumber\\
    &\equiv e^{n\epsilon t}\sum_{\mathclap{\;\;\;\{I_i\}_{i=1}^n}}\int\prod_{i=1}^n\Big\{\mathrm{d}t_if_{I_i}(t_i)\Big\}\chi_{AB_{I_1}\cdots B_{I_n}}^{(n)}(t-t_1,\cdots,t-t_n)e^{-\epsilon\sum_{i=1}^n(t-t_i)},
\end{flalign}
where $\sum_{\{I_i\}_{i=1}^n}=\sum_{I_1}\cdots\sum_{I_n}$.
Fourier transforming to frequency space, this yields
\begin{flalign}
\label{multi_perturbation_causal_response}
    \mathrm{tr}(A\rho_n(t))=\int{\mathrm{d}\omega}\sum_{\mathclap{\;\;\;\{I_i\}_{i=1}^n}}\int\Big\{\prod_{i=1}^n\frac{\mathrm{d}\omega_i}{2\pi}e^{-i\omega_i^+t}f_{I_i}(\omega_i)\Big\}\delta\Big(\omega-\sum_{i=1}^n\omega_i\Big)\chi_{AB_{I_1}\cdots B_{I_n}}^{(n)}(\omega_1^+,\cdots,\omega_n^+).
\end{flalign}
Examining Eq.~\eqref{multi_perturbation_causal_response}, we see that all terms besides $\chi_{AB_{I_1}\cdots B_{I_n}}^{(n)}(\omega_1^+,\cdots,\omega_n^+)$ are symmetric under the simultaneous exchange $(I_i,\omega_i)\leftrightarrow(I_j,\omega_j)$. 
Under a relabeling of dummy indices, this implies that the causal response function inherits the same symmetry, i.e.
\begin{equation}
    \chi_{AB_{I_1}\cdots B_{I_i}\cdots B_{I_j}\cdots B_{I_n}}^{(n)}(\omega_1,\cdots,\omega_i,\cdots,\omega_j,\cdots,\omega_n)=\chi_{AB_{I_1}\cdots B_{I_j}\cdots B_{I_i}\cdots B_{I_n}}^{(n)}(\omega_1,\cdots,\omega_j,\cdots,\omega_i,\cdots,\omega_n)\nonumber.
\end{equation}
We now plug Eq. \eqref{multi_perturbation_causal_response} into the equation of motion of $\mathrm{tr}(A\rho_n(t))$,
\begin{flalign}
\label{trace_EOM_multiple_perturbations}
    \mathrm{tr}(A\dot\rho_n(t))&=-i\mathrm{tr}(A[H_0,\rho_n(t)])-ie^{\epsilon t}\sum_If_I(t)\mathrm{tr}(A[B_I,\rho_{n-1}(t)])\nonumber\\
    &=-i\mathrm{tr}([A,H_0]\rho_n(t))-ie^{\epsilon t}\sum_If_I(t)\mathrm{tr}([A,B_I]\rho_{n-1}(t)).
\end{flalign}
The left hand side and the first term on the right hand side of Eq. \eqref{trace_EOM_multiple_perturbations} can be written in frequency space using Eq.~\eqref{multi_perturbation_causal_response} as
\begin{flalign}
    &\mathrm{tr}(A\dot\rho_n(t))\nonumber\\&=\int{\mathrm{d}\omega}\sum_{\mathclap{\;\;\;\{I_i\}_{i=1}^n}}\int\Big\{\prod_{i=1}^n\frac{\mathrm{d}\omega_i}{2\pi}e^{-i\omega_i^+t}f_{I_i}(\omega_i)\Big\}\Big(-i\sum_{i=1}^n\omega_i^+\Big)\chi_{AB_{I_1}\cdots B_{I_n}}^{(n)}(\omega_1^+,\cdots,\omega_n^+)\delta\Big(\omega-\sum_{i=1}^n\omega_i\Big),
\end{flalign}
and,
\begin{flalign}
    &\mathrm{tr}([A,H_0]\rho_n(t))\nonumber\\&=\int{\mathrm{d}\omega}\sum_{\mathclap{\;\;\;\{I_i\}_{i=1}^n}}\int\Big\{\prod_{i=1}^n\frac{\mathrm{d}\omega_i}{2\pi}e^{-i\omega_i^+t}f_{I_i}(\omega_i)\Big\}\chi_{[A,H_0]B_{I_1}\cdots B_{I_n}}^{(n)}(\omega_1^+,\cdots,\omega_n^+)\delta\Big(\omega-\sum_{i=1}^n\omega_i\Big).
\end{flalign}
The second term on the right hand side of Eq. \eqref{trace_EOM_multiple_perturbations} is similarly
\begin{flalign}\label{eq:multiple_eom_2nd_term}
    &e^{\epsilon t}\sum_{I_1}f_{I_1}(t)\mathrm{tr}([A,B_{I_1}]\rho_{n-1}(t))\nonumber\\&=e^{\epsilon t}\sum_{I_1}f_{I_1}(t)\int{\mathrm{d}\omega}\sum_{\mathclap{\;\;\;\{I_i\}_{i=2}^n}}\int\Big\{\prod_{i=2}^n\frac{\mathrm{d}\omega_i}{2\pi}e^{-i\omega_i^+t}f_{I_i}(\omega_i)\Big\}\chi_{[A,B_{I_1}]B_{I_2}\cdots B_{I_n}}^{(n)}(\omega_2^+,\cdots,\omega_n^+)\delta\Big(\omega-\sum_{i=2}^n\omega_i\Big)\nonumber\\&=\int\frac{\mathrm{d}\omega_1}{2\pi}e^{-i\omega_1^+t}\sum_{I_1}f_{I_1}(\omega_1)\nonumber\\&\times\int{\mathrm{d}\omega}\sum_{\mathclap{\;\;\;\{I_i\}_{i=2}^n}}\int\Big\{\prod_{i=2}^n\frac{\mathrm{d}\omega_i}{2\pi}e^{-i\omega_i^+t}f_{I_i}(\omega_i)\Big\}\chi_{[A,B_{I_1}]B_{I_2}\cdots B_{I_n}}^{(n)}(\omega_2^+,\cdots,\omega_n^+)\delta\Big(\omega-\sum_{i=2}^n\omega_i\Big)\nonumber\\&=\int\mathrm{d}\omega'\sum_{\mathclap{\;\;\;\{I_i\}_{i=1}^n}}\int\Big\{\prod_{i=1}^n\frac{\mathrm{d}\omega_i}{2\pi}e^{-i\omega_i^+t}f_{I_i}(\omega_i)\Big\}\chi_{[A,B_{I_1}]B_{I_2}\cdots B_{I_n}}^{(n)}(\omega_2^+,\cdots,\omega_n^+)\delta\Big(\omega'-\sum_{i=1}^n\omega_i\Big),
\end{flalign}
where in the last line we shifted $\omega$ to $\omega'=\omega+\omega_1$.
Again, using the explicit symmetry of the summation and integration measure under $(I_1,\omega_1)\leftrightarrow (I_{i\neq 1},\omega_{i\neq 1})$, we can rewrite Eq.~\eqref{eq:multiple_eom_2nd_term} as
\begin{flalign}
    &e^{\epsilon t}\sum_{I_1}f_{I_1}(t)\mathrm{tr}([A,B_{I_1}]\rho_{n-1}(t))\nonumber\\&=\int\mathrm{d}\omega'\sum_{\mathclap{\;\;\;\{I_i\}_{i=1}^n}}\int\Big\{\prod_{i=1}^n\frac{\mathrm{d}\omega_i}{2\pi}e^{-i\omega_i^+t}f_{I_i}(\omega_i)\Big\}\nonumber\\&\times\frac{1}{n}\Big(\chi_{[A,B_{I_1}]B_{I_2}\cdots B_{I_n}}^{(n)}(\omega_2^+,\cdots,\omega_n^+)+(I_1,\omega_1\leftrightarrow I_{i\neq 1},\omega_{i\neq 1})\Big)\delta\Big(\omega'-\sum_{i=1}^n\omega_i\Big).
\end{flalign}
Therefore, for multiple perturbations we find that the response function satisfies
\begin{flalign}
    \label{characteristic_equation_multiple_perturbation}
    &\Big(\sum_{i=1}^n\omega_i^+\Big)\chi^{(n)}_{AB_{I_1}\cdots B_{I_n}}(\omega^+_1,\cdots,\omega^+_n)\nonumber\\&=\chi^{(n)}_{[A,H_0]B_{I_1}\cdots B_{I_n}}(\omega^+_1,\cdots,\omega^+_n)+\frac{1}{n}\Big\{\chi^{(n-1)}_{[A,B_{I_1}]B_{I_2}\cdots B_{I_n}}(\omega^+_2,\cdots,\omega^+_n)+(I_1,\omega_1\leftrightarrow I_{i\neq 1},\omega_{i\neq 1})\Big\}.
\end{flalign}{}
We again use the Lehmann representation to write (note that there is no implicit sum over the $I_i$),
\begin{flalign}
    \chi^{(n)}_{AB_{I_1}\cdots B_{I_n}}(\omega_1,\cdots,\omega_n)=\frac{1}{\mathcal{Z}}\sum_{i_1,\cdots,i_{n+1}}(B_{I_1})^{i_1}_{i_2}\cdots (B_{I_n})^{i_n}_{i_{n+1}}A^{i_{n+1}}_{i_1}f^{(I_1\cdots I_n)}_{i_1,\cdots,i_{n+1}}(\omega_1,\cdots,\omega_n).
\end{flalign}
Plugging this into Eq. \eqref{characteristic_equation_multiple_perturbation} arrive at the recurrence relation for the Lehmann representation of the causal response function,
\begin{flalign}
\label{multi_perturbation_rec1}
    &f^{(I_1\cdots I_n)}_{i_1,\cdots,i_{n+1}}(\omega_1,\cdots,\omega_n)\nonumber\\&=\frac{1}{n(\omega_t^+-\epsilon_{i_1i_{n+1}})}\Bigg\{f^{(I_2\cdots I_n)}_{i_2,\cdots,i_{n+1}}(\omega_2,\cdots,\omega_n)-f^{(I_2\cdots I_n)}_{i_1,\cdots,i_{n}}(\omega_2,\cdots,\omega_n)+(I_1,\omega_1\leftrightarrow I_{i\neq 1},\omega_{i\neq 1})\Bigg\}.
\end{flalign}

We now compare Eq. \eqref{multi_perturbation_rec1}, to the Lehmann representation for the time-ordered correlation function
\begin{equation}
    \bar{\chi}_{AB_{I_1}\cdots B_{I_n}}^{(n)}(\tau_1,\cdots,\tau_n)=\frac{1}{n!}\mathrm{tr}(\mathbb{T}A(0)B_{I_1}(-\tau_1)\cdots B_{I_n}(-\tau_n)\rho_0).
\end{equation}
In frequency space, this imaginary time ordered response function has the same symmetrization property as, $\chi^{(n)}_{AB_{I_1}\cdots B_{I_n}}(\omega_1,\cdots,\omega_n)$ due to the time ordering symbol.
The imaginary time correlation function satisfies
\begin{flalign}\label{eq:chain_rule_2}
    \sum_{i=1}^n\frac{\partial}{\partial\tau_i}\bar{\chi}^{(n)}_{AB_{I_1}\cdots B_{I_n}}(\tau_1,\cdots,\tau_n)=-\bar{\chi}^{(n)}_{[A,H_0]B_{I_1}\cdots B_{I_n}}(\tau_1,\cdots,\tau_n),
\end{flalign}
following the same logic as in Eq.~\eqref{chain_rule}. 
We can use Eq.~\eqref{eq:chain_rule_2} to Fourier transform the time-ordered correlation function to find
\begin{flalign}\label{eq:multiple_ft}
    &\int_0^{\beta}\mathrm{d}\tau_ke^{i\nu_k\tau_k}\frac{\partial}{\partial\tau_k}\bar{\chi}^{(n)}_{AB_{I_1}\cdots B_{I_n}}(\tau_1,\cdots,\tau_n)\nonumber\\&=-i\nu_k\bar{\chi}^{(n)}_{AB_{I_1}\cdots B_{I_n}}(\tau_1,\cdots,\nu_k,\cdots,\tau_n)\nonumber\\&+\bar{\chi}^{(n)}_{AB_{I_1}\cdots B_{I_n}}(\tau_1,\cdots,\beta,\cdots,\tau_n)-\bar{\chi}^{(n)}_{AB_{I_1}\cdots B_{I_n}}(\tau_1,\cdots,0,\cdots,\tau_n).
\end{flalign}
Using the time ordering symbol, we have,
\begin{flalign}
\label{eq:multiple_0}
    \bar{\chi}^{(n)}_{AB_{I_1}\cdots B_{I_n}}(\tau_1,\cdots,0,\cdots,\tau_n)&=\frac{1}{n!}\mathrm{tr}(A\mathbb{T}B_{I_1}(-\tau_1)\cdots B_{I_k}(0)\cdots B_{I_n}(-\tau_n)\rho_0)\nonumber\\&=\frac{1}{n!}\mathrm{tr}(AB_{I_k}\mathbb{T}B_{I_1}(-\tau_1)\cdots \xcancel{B_{I_k}(\tau_k)}\cdots B_{I_n}(-\tau_n)\rho_0)\nonumber\\&=\frac{1}{n}\bar{\chi}^{(n-1)}_{(AB_{I_k})B_{I_1}\cdots\xcancel{B_{I_k}}\cdots B_{I_n}}(\tau_1,\cdots,\xcancel{\tau_k},\cdots,\tau_n)
\end{flalign}
and similarly,
\begin{flalign}\label{eq:multiple_beta}
    \bar{\chi}^{(n)}_{AB_{I_1}\cdots B_{I_n}}(\tau_1,\cdots,\beta,\cdots,\tau_n)&=\frac{1}{n!}\mathrm{tr}(A\mathbb{T}B_{I_1}(-\tau_1)\cdots B_{I_k}(-\beta)\cdots B_{I_n}(-\tau_n)\rho_0)\nonumber\\&=\frac{1}{n!}\mathrm{tr}(A\mathbb{T}B_{I_1}(-\tau_1)\cdots \xcancel{B_{I_k}(-\beta)}\cdots B_{I_n}(-\tau_n)\rho_0 B_{I_k}\rho^{-1}_0\rho_0)\nonumber\\&=\frac{1}{n!}\mathrm{tr}(B_{I_k}A\mathbb{T}B_{I_1}(-\tau_1)\cdots\xcancel{B_{I_k}(-\tau_k)}\cdots B_{I_n}(-\tau_n)\rho_0)\nonumber\\
    &=\frac{1}{n}\bar{\chi}^{(n-1)}_{(B_{I_k}A)B_{I_1}\cdots\xcancel{B_{I_k}}\cdots B_{I_n}}(\tau_1,\cdots,\xcancel{\tau_k},\cdots,\tau_n).
\end{flalign}
Combining Eqs.~\eqref{eq:multiple_ft}--\eqref{eq:multiple_beta}, we see that the characteristic equation for the Matsubara Green's function is
\begin{flalign}\label{eq:matsubara_recurrence_multiple}
    &\Big(-i\sum_{i=1}^n\nu_i\Big)\bar{\chi}^{(n)}_{AB_{I_1}\cdots B_{I_n}}(\nu_1,\cdots,\nu_n)\nonumber\\&=-\bar{\chi}^{(n)}_{[A,H_0]B_{I_1}\cdots B_{I_n}}(\nu_1,\cdots,\nu_n)+\frac{1}{n}\Big\{\chi^{(n-1)}_{[A,B_{I_1}]B_{I_2}\cdots B_{I_n}}(\nu_2,\cdots,\nu_n)+(I_1,\nu_1\leftrightarrow I_{i\neq 1},\nu_{i\neq 1})\Big\}.
\end{flalign}
We use the Lehmann representation to write (with no implied sum over $I_i$)
\begin{flalign}
    \bar{\chi}^{(n)}_{AB_{I_1}\cdots B_{I_n}}(\nu_1,\cdots,\nu_n)=\frac{1}{\mathcal{Z}}\sum_{i_1,\cdots,i_{n+1}}(B_{I_1})^{i_1}_{i_2}\cdots (B_{I_n})^{i_n}_{i_{n+1}}A^{i_{n+1}}_{i_1}g^{(I_1\cdots I_n)}_{i_1,\cdots,i_{n+1}}(\nu_1,\cdots,\nu_n).
\end{flalign}
When inserted into the characteristic equation Eq.~\eqref{eq:matsubara_recurrence_multiple} we arrive at the recurrence relation
\begin{flalign}
\label{multi_perturbation_rec2}
    &g^{(I_1\cdots I_n)}_{i_1,\cdots,i_{n+1}}(\nu_1,\cdots,\nu_n)\nonumber\\&=\frac{-1}{n(i\nu_t^+-\epsilon_{i_1i_{n+1}})}\Bigg\{g^{(I_2\cdots I_n)}_{i_2,\cdots,i_{n+1}}(\nu_2,\cdots,\nu_n)-g^{(I_2\cdots I_n)}_{i_1,\cdots,i_{n}}(\nu_2,\cdots,\nu_n)+(I_1,\nu_1\leftrightarrow I_{i\neq 1},\nu_{i\neq 1})\Bigg\}.
\end{flalign}
Comparing Eq.~\eqref{multi_perturbation_rec2} with Eq. \eqref{multi_perturbation_rec1}, we see that the Matsubara Green's function can be related to the causal response function using induction if we can prove the $n=1$ case.
For $n=1$, the causal response function,
\begin{equation}
    f^{I}_{i_1,i_2}(\omega)=\frac{e^{-\beta\epsilon_{i_2}}-e^{-\beta\epsilon_{i_1}}}{(\omega^+-\epsilon_{i_1i_2})},
\end{equation}
is independent of $I$.
We compare this to,
\begin{flalign}
    \int_0^{\beta}\mathrm{d}\tau e^{i\nu\tau}\mathrm{tr}(AB_I(-\tau)\rho_0)&=\frac{1}{\mathcal{Z}}\sum_{i_1,i_2}(B_I)^{i_1}_{i_2}A^{i_2}_{i_1}e^{-\beta\epsilon_{i_2}}\frac{e^{\beta\epsilon_{i_2i_1}}-1}{i\nu+\epsilon_{i_2i_1}}\nonumber\\&=-\frac{1}{\mathcal{Z}}\sum_{i_1,i_2}(B_I)^{i_1}_{i_2}A^{i_2}_{i_1}\frac{e^{-\beta\epsilon_{i_2}}-e^{-\beta\epsilon_{i_1}}}{i\nu-\epsilon_{i_1i_2}}\nonumber\\&=-\frac{1}{\mathcal{Z}}\sum_{i_1,i_2}(B_I)^{i_1}_{i_2}A^{i_2}_{i_1}f^{I}_{i_1,i_2}(\omega^+=i\nu)\nonumber.
\end{flalign}
Therefore, we have $f^I_{i_1,i_2}(\omega^+)=-g^I_{i_1,i_2}(i\nu=\omega^+)$, which proves that for all $n$,
\begin{equation}
    \chi^{(n)}_{AB_{I_1}\cdots B_{I_n}}(\omega_1^+,\cdots,\omega_n^+)=(-1)^n\bar{\chi}^{(n)}_{AB_{I_1}\cdots B_{I_n}}(i\nu_1=\omega_1^+,\cdots,i\nu_n=\omega_n^+).
\end{equation}

\section{Spectral density representation of Response and Matsubara Functions}
\label{Spectral density representation of Response and Matsubara Functions}
Our proof of equivalence between causal response functions and Matsubara functions relies on our ability to unambiguously analytically continue the causal response function to the upper half frequency plane. 
To carry out this analytic continuation, we relied on the Lehmann representation of the Green's functions, where all singularities were given by a sum of simple poles. 
However, this implicitly assumes that the many-body energies of the system under consideration are discrete. 
While this is true for any finite-sized system, care must be taken to extend our present argument to the thermodynamic limit.

In order to circumvent this difficulty, we will now present an alternative approach based on the spectral density representation of the response function. 
We will proceed by first deriving the spectral density representation of the response function $\chi^{(n)}_{AB_{I_1}\cdots B_{I_n}}$, and show explicitly that it satisfies the equations of motion Eq.~\eqref{characteristic_equation_multiple_perturbation}. 
Using these facts, we will then show how to perform the analytic continuation to obtain the Matsubara functions.

To begin, we first solve Eq.~\eqref{eq:rhoEOM_multiple} exactly using iterated commutators. 
We note that $\rho(t)$ can be formally expressed using the time-evolution operator as
\begin{equation}
\rho(t)=U(t)\rho_0U^\dag(t),
\end{equation}
where the time-evolution operator is given as a time-ordered exponential
\begin{equation}
U(t)=\mathbb{T}\exp\left(-i\int_{-\infty}^tdt'H(t')\right)
\end{equation}
Expanding $U(t)$ as a Dyson series and retaining the terms that are $n$-th order in $f_I(t)$ yields~\cite{kubo1957statistical}
\begin{equation}\label{eq:rho_n_explicit}
\rho_n(t)= (-i)^n\int_{-\infty}^tdt_1\int_{-\infty}^{t_1}dt_2\dots\int_{-\infty}^{t_{n-1}}dt_n \left(\prod_{i=1}^{n}f_{I_{i}}(t_i)e^{\epsilon t_i}\right)\underbrace{\left[B_{I_1}(t_1-t),\left[B_{I_2}(t_2-t),\left[\dots \left[B_{I_n}(t_n-t),\rho_0\right]\right]\right]\right]}_{\text{$n$ times}},
\end{equation}
where the time evolution of $B_{I_i}(t)$ is defined in the in interaction picture,
\begin{equation}
B_{I_i}(t) \equiv e^{iH_0t}B_{I_i}e^{-iH_0t}.
\end{equation}
That Eq.~\eqref{eq:rho_n_explicit} satisfies Eq.~\eqref{eq:rhoEOM_multiple} can be verified by directly differentiating with respect to $t$: the first term on the right-hand-side of Eq.~\eqref{eq:rhoEOM_multiple} comes from differentiating the $t$-dependence of the $B_{I_i}$'s, while the second term comes from the derivative acting on the upper limit of the $t_1$ integration.

Next, we can substitute Eq.~\eqref{eq:rho_n_explicit} into our defining equation Eq.~\eqref{eq:responsedef_multiple_t} for the $n$-th order response function. 
Doing so, we find
\begin{align}
\mathrm{tr}(A\rho_n(t)) &= (-i)^n\int_{-\infty}^tdt_1\int_{-\infty}^{t_1}dt_2\dots\int_{-\infty}^{t_{n-1}}dt_n \left(\prod_{i=1}^{n}f_{I_{i}}(t_i)e^{\epsilon t_i}\right)\mathrm{tr}\left(A\underbrace{\left[B_{I_1}(t_1-t),\left[B_{I_2}(t_2-t),\left[\dots \left[B_{I_n}(t_n-t),\rho_0\right]\right]\right]\right]}_{\text{$n$ times}}\right) \nonumber \\
&=(-i)^n e^{n\epsilon t}\int_{-\infty}^t\int_{-\infty}^{t_1}\dots\int_{-\infty}^{t_{n-1}}\left(\prod_{i=1}^{n}dt_i f_{I_{i}}(t_i)e^{\epsilon (t-t_i)}\right)\langle\underbrace{\left[\left[\left[\left[A(t),B_{I_1}(t_1)\right],B_{I_2}(t_2)\right],\dots\right],B_{I_n}(t_n)\right]}_{\text{$n$ times}}\rangle_0.
\end{align}
Comparing with the right-hand side of Eq.~\eqref{eq:responsedef_multiple_t}, we can use time-translation invariance of the unperturbed expectation value to identify the \emph{unsymmetrized} response function
\begin{equation}
\tilde{\chi}_{AB_{I_1}\cdots B_{I_n}}^{(n)}(t-t_1,\cdots,t-t_n) = (-i)^n\Theta(t-t_1)\Theta(t_1-t_2)\dots\Theta(t_{n-1}-t_n)\langle\underbrace{\left[\left[\left[\left[A(t),B_{I_1}(t_1)\right],B_{I_2}(t_2)\right],\dots\right],B_{I_n}(t_n)\right]}_{\text{$n$ times}}\rangle_0e^{\epsilon(\sum_i t_i-t)},
\end{equation}
in terms of which the full symmetrized response function is given by 
\begin{equation}
\chi_{AB_{I_1}\cdots B_{I_n}}^{(n)}(t-t_1,\cdots,t-t_n)=\frac{1}{n!}\sum_{\sigma}\tilde{\chi}_{AB_{I_{\sigma(1)}}\cdots B_{I_{\sigma(n)}}}^{(n)}(t-t_{\sigma(1)},\cdots,t-t_{\sigma(n)}),
\end{equation}
where the summation is over the $n!$ permutations $\sigma$ of $n$ elements.

Using the convolution theorem, we can take the Fourier transform of Eq.~\eqref{eq:responsedef_multiple_t} to obtain Eq.~\eqref{multi_perturbation_causal_response}. 
Using the change of variables $v_i = t_{i-1}-t_i, v_1=t-t_1$ we can write
\begin{equation}
\chi_{AB_{I_1}\cdots B_{I_n}}^{(n)}(\omega_1,\dots\omega_n)=\frac{1}{n!}\sum_{\sigma}\tilde{\chi}_{AB_{I_{\sigma(1)}}\cdots B_{I_{\sigma(n)}}}^{(n)}(\omega_{\sigma(1)},\cdots,\omega_{\sigma(n)})
\end{equation}
with
\begin{equation}\label{eq:nth_order_kubo}
\tilde{\chi}_{AB_{I_{1}}\cdots B_{I_{n}}}^{(n)}(\omega_{1},\cdots,\omega_{n}) = (-i)^n\int \prod_{i=1}^n\left(dv_i\Theta(v_i)e^{iv_i\sum_{j=i}^n\omega_j^+}\right)\langle\underbrace{\left[\left[\left[\left[A(0),B_{I_1}(-v_1)\right],B_{I_2}(-v_1-v_2)\right],\dots\right],B_{I_n}(-\sum_{i=1}^n v_i)\right]}_{\text{$n$ times}}\rangle_0
\end{equation}
Eq.~\eqref{eq:nth_order_kubo} is nothing other than the $n$-th order generalization of the Kubo formula, written in a particularly convenient set of coordinates. 
We can use the spectral representation of the step function
\begin{equation}
\Theta(v)e^{i\omega^+ v} = \frac{1}{2\pi i}\int d\alpha \frac{e^{i\alpha v}}{\alpha-\omega^+}
\end{equation}
to rewrite Eq.~\eqref{eq:nth_order_kubo} as
\begin{equation}\label{eq:tilde_chi_spectral}
\tilde{\chi}_{AB_{I_{1}}\cdots B_{I_{n}}}^{(n)}(\omega_{1},\cdots,\omega_{n}) = \left(\frac{-1}{\pi}\right)^n\int\left(\prod_{i=1}^n \frac{d\alpha_i}{\alpha_i -\sum_{j=i}^n\omega_j^+}\right)\eta^{(n)}_{AB_{I_{1}}\cdots B_{I_{n}}} (\alpha_1,\dots,\alpha_n)
\end{equation}
where $\eta^{(n)}_{AB_{I_{1}}\cdots B_{I_{n}}}(\alpha_1,\dots\alpha_n)$ is the spectral density defined as
\begin{equation}\label{eq:specdensitydef}
\eta^{(n)}_{AB_{I_{1}}\cdots B_{I_{n}}}(\alpha_1,\dots\alpha_n)=\frac{1}{2^n}\int \prod_{i=1}^n\left(dv_ie^{iv_i\alpha_i}\right)\langle\underbrace{\left[\left[\left[\left[A(0),B_{I_1}(-v_1)\right],B_{I_2}(-v_1-v_2)\right],\dots\right],B_{I_n}(-\sum_{i=1}^n v_i)\right]}_{\text{$n$ times}}\rangle_0.
\end{equation}
Note that if we insert complete sets of states into Eq.~\eqref{eq:specdensitydef}, the time evolution of each $B_I$ operator gives a phase. 
We can then carry out each integral over $v_i$ to obtain a product of delta functions. 
We thus see that the spectral density in the Lehmann representation is a (possibly continuous) sum of delta functions.

Explicitly symmetrizing Eq.~\eqref{eq:tilde_chi_spectral}, we the spectral representation of the causal response function,
 \begin{equation}
\chi_{AB_{I_1}\cdots B_{I_n}}^{(n)}(\omega_1,\dots\omega_n) =\frac{(-1)^n}{\pi^n n!}\sum_\sigma\int\left(\prod_{i=1}^n \frac{d\alpha_i}{\alpha_i -\sum_{j=i}^n\omega_{\sigma(j)}^+}\right)\eta^{(n)}_{AB_{I_{\sigma(1)}}\cdots B_{I_{\sigma(n)}}} (\alpha_1,\dots,\alpha_n).
 \end{equation}
 Each term in the sum over permutations $\sigma$ is a meromorphic function of all the $\omega_i$ that is analytic in each upper half plane. 
 Additionally, the location of the poles are independent of the spectral density, so analytic continuation to the complex $\omega_i$ planes can be carried out independently of the spectral density. 
 We can thus define the analytically-continued function
 \begin{equation}\label{eq:Xdef_appendix}
 X_{AB_{I_1}\cdots B_{I_n}}^{(n)}(z_1,\dots,z_n) = \frac{(-1)^n}{\pi^n n!}\sum_\sigma\int\left(\prod_i \frac{d\alpha_i}{\alpha_i -\sum_{j=i}^nz_{\sigma(j)}}\right)\eta^{(n)}_{AB_{I_{\sigma(1)}}\cdots B_{I_{\sigma(n)}}} (\alpha_1,\dots,\alpha_n),
 \end{equation}
 satisfying
 \begin{equation}\label{eq:chifromX}
 X_{AB_{I_1}\cdots B_{I_n}}^{(n)}(z_1=\omega_1^+,\dots,z_n=\omega_n^+) = \chi_{AB_{I_1}\cdots B_{I_n}}^{(n)}(\omega_1,\dots\omega_n).
 \end{equation}
 What is more, using the explicit expression Eq.~\eqref{eq:specdensitydef}, we have that
 \begin{align}\label{eq:Xsimplification}
\sum_{i=1}^{n}z_iX_{AB_{I_1}\cdots B_{I_n}}^{(n)}(z_1,\dots,z_n) &= \frac{(-1)^n}{\pi^n n!}\sum_\sigma\int\left(\prod_i \frac{d\alpha_i \sum_{j=1}^n z_j} {\alpha_i -\sum_{j=i}^nz_{\sigma(j)}}\right)\eta^{(n)}_{AB_{I_{\sigma(1)}}\cdots B_{I_{\sigma(n)}}} (\alpha_1,\dots,\alpha_n) \nonumber \\
&=\frac{(-1)^{n-1}}{\pi^n n!}\sum_\sigma\int\left(\prod_{i=2}^n \frac{d\alpha_i} {\alpha_i -\sum_{j=i}^nz_{\sigma(j)}}\right)d\alpha_1\left(1-\frac{\alpha_1}{\alpha_1-\sum_{j=1}^{n}z_j}\right)\eta^{(n)}_{AB_{I_{\sigma(1)}}\cdots B_{I_{\sigma(n)}}} (\alpha_1,\dots,\alpha_n).
 \end{align}
 To further simplify Eq.~\eqref{eq:Xsimplification}, we use the following identities that follow from integrating the explicit expression Eq.~\eqref{eq:specdensitydef} for $\eta^{(n)}_{AB_{I_{1}}\cdots B_{I_{n}}} (\alpha_1,\dots,\alpha_n)$:
 \begin{align}
\int d\alpha_1 \eta^{(n)}_{AB_{I_{1}}\cdots B_{I_{n}}} (\alpha_1,\dots,\alpha_n) &= \pi\eta^{(n-1)}_{[A,B_{I_{1}}]B_{I_2}\cdots B_{I_{n}}}(\alpha_2,\dots,\alpha_n)\label{eq:specdensityidentity1} \\
\alpha_1\eta^{(n)}_{AB_{I_{1}}\cdots B_{I_{n}}} (\alpha_1,\dots,\alpha_n) &= \eta^{(n)}_{[A,H_0]B_{I_{1}}\cdots B_{I_{n}}} (\alpha_1,\dots,\alpha_n)\label{specdensityidentity2}
\end{align}
Inserting Eqs.~\eqref{eq:specdensityidentity1} and \eqref{specdensityidentity2} into Eq.~\eqref{eq:Xsimplification} and using Eq.~\eqref{eq:Xdef_appendix}, we have
\begin{align}\label{eq:generalrecurrence_appendix}
\sum_{i=1}^{n}z_i&X_{AB_{I_1}\cdots B_{I_n}}^{(n)}(z_1,\dots,z_n) \nonumber \\
&= X_{[A,H_0]B_{I_1}\cdots B_{I_n}}^{(n)}(z_1,\dots,z_n) + \frac{1}{n}\left\{ X^{(n-1)}_{[A,B_{I_{1}}]B_{I_{2}}\cdots B_{I_{n}}}(z_{2},\dots,z_{n}) + (I_1,z_1)\leftrightarrow (I_{i\neq 1},z_{i\neq 1})\right\},
\end{align}
where we made use of the fact that the sum over permutations of $n$ elements $\{I_1,z_1\},\dots,\{I_n,z_n\}$ splits into a sum over permutations of the $n-1$ elements $\{I_2,z_2\},\dots,\{I_n,z_n\}$, followed by explicitly symmetrizing under $(I_1,z_1)\leftrightarrow (I_{i\neq 1},z_{i\neq 1})$.
For $z_i\rightarrow \omega_i^+$, Eq.~\eqref{eq:generalrecurrence_appendix} recovers the recurrence relation Eq.~\eqref{characteristic_equation_multiple_perturbation} for the nonlinear response function, as required by Eq.~\eqref{eq:chifromX}. 
Importantly, however, letting $z_i\rightarrow i\nu_i$ in the upper half plane, we see using Eq.~\eqref{eq:generalrecurrence_appendix} that $(-1)^{n}X_{AB_{I_1}\cdots B_{I_n}}^{(n)}(i\nu_1,\dots,i\nu_n)$ satisfies the recurrence relation Eq.~\eqref{eq:matsubara_recurrence_multiple}. 
We also know from linear response theory that for $n=1$, $-X_{AB_{I_1}}^{(1)}(i\nu_1)$ is the Matsubara two-point function. 
Thus, by induction, $(-1)^{n}X_{AB_{I_1}\cdots B_{I_n}}^{(n)}(i\nu_1,\dots,i\nu_n)$ is the Matsubara $n$-point function. 
This is our desired result.

\twocolumngrid
\bibliography{refs.bib} 

\end{document}